\documentclass[iop, superscriptaddress, reprint, graphicx]{revtex4-1}

\usepackage{color}

\newcommand{\lp}{\left(}
\newcommand{\rp}{\right)}

\newcommand{\lsb}{\left[}
\newcommand{\rsb}{\right]}

\newcommand{\refeq}[1]{Equation (\ref{#1})}
\newcommand{\refsec}[1]{Section \ref{#1}}
\newcommand{\refapp}[1]{Appendix \ref{#1}}
\newcommand{\reffig}[1]{Figure \ref{#1}}
\newcommand{\refmultifig}[2]{Figures \ref{#1} and \ref{#2}}
\newcommand{\reftable}[1]{Table \ref{#1}}

\newcommand{\correction}[1]{\textcolor{black}{#1}}

\newcommand{\KAISTNQe}{Department of Nuclear and Quantum Engineering, KAIST, Daejeon 34141, Republic of Korea}

\newcommand{\MaxPlanckGreifswald}{Max-Planck-Institut f$\ddot{u}$r Plasmaphysik, Teilinstitut Greifswald, D-17491 Greifswald, Germany}

\newcommand{\NFRI}{National Fusion Research Institute, Daejeon 34133, Republic of Korea}

\newcommand{\MOBIIS}{Mobiis Co., Ltd., Seongnam-si, Gyeonggi-do 13486, Republic of Korea}

\newcommand{\smjoung}{\author{Semin~Joung}\email[]{smjoung@kaist.ac.kr}\affiliation{\KAISTNQe}}
\newcommand{\jwkim}{\author{Jaewook~Kim}\affiliation{\KAISTNQe}}

\newcommand{\ycghim}{\author{Y.-c.~Ghim}\email[]{ycghim@kaist.ac.kr}\affiliation{\KAISTNQe}}
\newcommand{\shkwak}{\author{Sehyun~Kwak}\affiliation{\KAISTNQe}\affiliation{\MaxPlanckGreifswald}}

\newcommand{\hshan}{\author{H.S.~Han}\affiliation{\NFRI}}
\newcommand{\hskim}{\author{H.S.~Kim}\affiliation{\NFRI}}
\newcommand{\jgbak}{\author{J.G.~Bak}\affiliation{\NFRI}}
\newcommand{\sglee}{\author{S.G.~Lee}\affiliation{\NFRI}}
\newcommand{\ghlee}{\author{Geunho~Lee}\affiliation{\MOBIIS}}
\newcommand{\dhkwon}{\author{Daeho~Kwon}\affiliation{\KAISTNQe}\affiliation{\MOBIIS}}

\expandafter\let\csname equation*\endcsname\relax
\expandafter\let\csname endequation*\endcsname\relax
\usepackage[toc,page]{appendix}
\usepackage{graphicx}
\usepackage{amstext}
\usepackage{amssymb}
\usepackage{amsmath}
\usepackage{booktabs}
\usepackage{array}
\usepackage{multirow}
\usepackage{amsbsy}
\usepackage{hyperref}
\hypersetup{
    colorlinks=true,
    linkcolor=blue,
    citecolor=blue,
}
\makeatletter
\newcommand*{\rom}[1]{\expandafter\@slowromancap\romannumeral #1@}
\makeatother

\begin{document}

\title{Deep neural network Grad-Shafranov solver constrained with measured magnetic signals}

\smjoung
\jwkim
\shkwak
\jgbak
\sglee
\hshan
\hskim
\ghlee
\dhkwon
\ycghim

\date{\today}
\begin{abstract}	
A neural network solving Grad-Shafranov equation constrained with measured magnetic signals to reconstruct magnetic equilibria in real time is developed. Database created to optimize the neural network's free parameters contain off-line EFIT results as the output of the network from $1,118$ KSTAR experimental discharges of two different campaigns. Input data to the network constitute magnetic signals measured by a Rogowski coil (plasma current), magnetic pick-up coils (normal and tangential components of magnetic fields) and flux loops (poloidal magnetic fluxes). The developed neural networks fully reconstruct not only the poloidal flux function $\psi\lp R, Z\rp$ but also the toroidal current density function $j_\phi\lp R, Z\rp$ with the off-line EFIT quality. To preserve robustness of the networks against a few missing input data, an imputation scheme is utilized to eliminate the required additional training sets with large number of possible combinations of the missing inputs.
\end{abstract}

\maketitle

%
\vspace{2pc}
\noindent{\it Keywords}: Neural network, Grad-Shafranov equation, EFIT, poloidal flux, toroidal current, imputation, KSTAR
%
%
%
%

\section{Introduction} \label{S1:intro}

Magnetic equilibrium is one of the most important information to understand the basic behavior of plasmas in magnetically confined plasmas, and the off-line EFIT \cite{Lao:1985hn} code has been extensively used to reconstruct such equilibria in tokamaks.  Its fundamentals are basically finding a solution to an ideal magnetohydrodynamic equilibrium with toroidal axisymmetry, known as the Grad-Shafranov (GS) equation \cite{Freidberg:1987}:
\begin{equation}
\begin{split} \label{eq:gseq}
\Delta^*\psi &\equiv \lp R\frac{\partial}{\partial R} \frac{1}{R} \frac{\partial}{\partial R} + \frac{\partial^2}{\partial Z^2} \rp \psi \\
& =  -\mu_{0}R j_\phi \\
& = -\mu_0 R^2 \frac{d p(\psi)}{d \psi} - F(\psi)\frac{d F(\psi)}{d \psi},
\end{split}
\end{equation}
where $\psi=\psi\lp R, Z\rp$ is the poloidal flux function, $j_\phi=j_\phi\lp R, Z\rp$ the toroidal current density function, $p(\psi)$ the plasma pressure. $F(\psi)$ is related to the net poloidal current. Here, $R$, $\phi$ and $Z$ denote the usual cylindrical coordinate system.  As the $\Delta^*$ is a two-dimensional nonlinear partial differential operator, the off-line EFIT \cite{Lao:1985hn} finds a solution with many numerical iterations and has been implemented in many tokamaks such as D\rom{3}-D \cite{Lao:2005kd}, JET \cite{OBrien:1992gm}, NSTX \cite{Sabbagh:2001hh}, EAST \cite{Jinping:2009jn} and KSTAR \cite{Park:2011go} to name some as examples.

With an aim of real-time control of tokamak plasmas, real-time EFIT (rt-EFIT) \cite{Ferron:2002fw} code is developed to provide a magnetic equilibrium fast enough whose results are different from the off-line EFIT results. As pulse lengths of tokamak discharges become longer \cite{VanHoutte:1993, Ekedahl_2010, Itoh_1999, Zushi_2003, Saoutic_2002, Park:2019, Wan:2019}, demand on more elaborate plasma control is ever increased. Furthermore, some of the ITER relevant issues such as ELM (edge localized mode) suppression with RMP (resonant magnetic perturbation) coils \cite{Park_NP:2018} and the detached plasma scenarios \cite{Reimold:2015, Jaervinen:2016} require sophisticated plasma controls, meaning that the more accurate magnetic equilibria we have in real time, the better performance we can achieve.

There has been an attempt to satisfy such a requirement of acquiring a more accurate, i.e., closer to the off-line EFIT results compared to the rt-EFIT results, magnetic equilibrium in real-time using graphics processing units (GPUs) \cite{Yue:2013cj} by parallelizing equilibrium reconstruction algorithms. The GPU based EFIT (P-EFIT)  \cite{Yue:2013cj} enabled one to calculate a well-converged equilibrium in much less time; however, the benchmark test showed similar results to the rt-EFIT rather than the off-line results \cite{Huang:2016gz}.

Thus, we propose a reconstruction algorithm based on a neural network that satisfies the GS equation as well as the measured magnetic signals to obtain accurate magnetic equilibrium in real time. We note that usage of neural networks in fusion community is increasing rapidly, and examples are radiated power estimation \cite{Barana:2002RSI}, identifying instabilities \cite{Murari:2013cm}, estimating neutral beam effects \cite{Boyer_2019}, classifying confinement regimes \cite{Murari:2012fl}, determination of scaling laws \cite{MURARI2010850, Gaudio_2014}, disruption prediction \cite{Tang:2019Nature, Cannas:NF2010, Pau:NF2019}, turbulent transport modelling \cite{Meneghini:2014ic, Meneghini:2017kp, Citrin:2015fj, Felici:2018db}, plasma tomography with the bolometer system \cite{Matos:2017kl, Ferreira:2018}, coil current prediction with the heat load pattern in W7-X \cite{Bockenhoff:2018hl}, filament detection on MAST-U \cite{Cannas:2019gr}, electron temperature profile estimation via SXR with Thomson scattering \cite{Clayton:2013dx} and equilibrium reconstruction \cite{Lister:1991gx, Coccorese:1994jt, Bishop:1994kr, Cacciola:2006bq, Jeon:2001bd, Wang:JFE2016} together with an equilibrium solver \cite{vanMilligen:1995dv}. Most of previous works on the equilibrium reconstruction with neural networks have paid attention to finding the poloidal beta, the plasma elongation, positions of the X-points and plasma boundaries, i.e., last closed flux surface, and gaps between plasmas and plasma facing components, rather than reconstructing the whole internal magnetic structures we present in this work.

The inputs to our developed neural networks consist of plasma current measured by a Rogowski coil, normal and tangential components of magnetic fields by magnetic pick-up coils, poloidal magnetic fluxes by flux loops and a position in $\lp R, Z\rp$ coordinate system, where $R$ is the major radius, and $Z$ is the height as shown in \reffig{fig:kstarConfig}. The output of the neural networks is a value of poloidal flux $\psi$ at the specified $\lp R, Z\rp$ position. To train and validate the neural networks, we have collected a total of $1,118$ KSTAR discharges from two consecutive campaigns, i.e., $2017$ and $2018$ campaigns. We, in fact, generate three separate neural networks which are NN$_\textrm{2017}$, NN$_\textrm{2018}$ and NN$_\textrm{2017, 2018}$ where subscripts indicate the year(s) of KSTAR campaign(s) that the training data sets are obtained from. Additional $163$ KSTAR discharges (from the same two campaigns) are collected to test the performance of the developed neural networks.

We train the neural networks with the KSTAR off-line EFIT results, and let them be \textit{accurate} magnetic equilibria. Note that disputing on whether the off-line EFIT results we use to train the networks are accurate or not is beyond the scope of this work. If we find more accurate EFIT results, e.g., MSE(Motional Stark Effect)-constrained EFIT or more sophisticated equilibrium reconstruction algorithms that can cope with current-hole configurations (current reversal in the core) \cite{Rodrigues:2005PRL, Rodrigues:2007PRL, Ludwig:2013NF}, then we can always re-train the networks with new sets of data as long as the networks follow the trained EFIT data with larger similarity than the rt-EFIT results do. This is because supervised neural networks are limited to follow the training data. Hence, as a part of the training sets we use the KSTAR off-line EFIT results  as possible examples of \textit{accurate} magnetic equilibria to corroborate our developed neural networks. 

To calculate the output data a typical neural network requires the same set of input data as it has been trained. Therefore, even a single missing input (out of input data set) can result in a flawed output  \cite{vanLint:2005cn}. Such a case can be circumvented by training the network with possible combinations of missing inputs. As a part of input data, we have $32$ normal and $36$ tangential magnetic fields measured by the magnetic pick-up coils. If we wish to cover a case with one missing input data, then we will need to repeat the whole training procedure with $68$ ($32+36$) different cases. If we wish to cover a case with two or three missing input data, then we will need additional $2,278$ and $50,116$ different cases to be trained on, respectively. This number becomes large rapidly, and it becomes formidable, if not impossible, to train the networks with reasonable computational resources. Since the magnetic pick-up coils are susceptible to damages, we have developed our networks to be capable of inferring a few missing signals of the magnetic pick-up coils in real-time by invoking an imputation scheme \cite{Joung:2018ju} based on Bayesian probability \cite{Sivia:2006} and Gaussian processes \cite{Rasmussen:2006}.

In addition to reconstructing \textit{accurate} magnetic equilibria in real-time, the expected improvements with our neural networks compared to the previous studies are at least fourfold: (1) the network is capable of providing whole internal magnetic topology, not limited to boundaries and locations of X-points and/or magnetic axis; (2) spatial resolution of reconstructed equilibria is arbitrarily adjustable within the first wall of KSTAR since $\lp R, Z\rp$ position is a part of the input data; (3) the required training time and computational resources for the networks are reduced by generating a coarse grid points also owing to $\lp R, Z\rp$ position being an input, and (4) the networks can handle a few missing signals of the magnetic pick-up coils using the imputation method.

We, first, present how the data are collected to train the neural networks and briefly discuss real-time preprocessing of the measured magnetic signals in \refsec{S2:collection}. For the readers who are interested in thorough description of the real-time preprocessing, \refapp{app:drfit_adjust} provides the details. Then, we explain the structure of our neural networks and how we train them in \refsec{S3:nn}. In \refsec{S4:trainefit}, we present the results of the developed neural network EFIT (nn-EFIT) in four aspects. First, we discuss how well the NN$_\textrm{2017, 2018}$ network reproduces the off-line EFIT results. Then, we make comparisons among the three networks, NN$_\textrm{2017}$, NN$_\textrm{2018}$ and NN$_\textrm{2017, 2018}$, by examining in-campaign and cross-campaign performance. Once the absolute performance qualities of the networks are established, we compare relative performance qualities between nn-EFIT and rt-EFIT. Finally, we show how the imputation method support the networks when there exist missing inputs. Our conclusions are presented in \refsec{S6:con}.

\section{Collection and real-time preprocessing of data}
\label{S2:collection}

\begin{figure}
    \centering
    \includegraphics[width=0.6\linewidth]{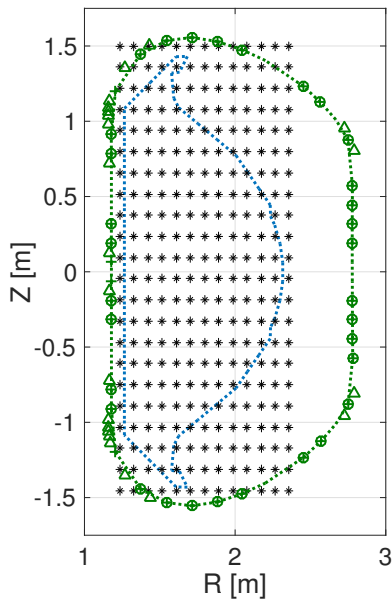}
    \caption{A poloidal cross-section of KSTAR with the first wall (blue dotted line). Green dotted line indicates a Rogowski coil measuring the plasma current ($I_\textrm{p}$). Green open circles and crosses depict locations of the magnetic pick-up coils measuring $32$ normal ($B_\textrm{n}$) and $36$ tangential ($B_\textrm{t}$) magnetic fields, respectively, whereas green triangles represent $22$ flux loops measuring poloidal magnetic fluxes ($\Psi_\textrm{FL}$). Black asterisks ($22\times 13$ spatial positions) show locations where we obtain the values of $\psi$ from the off-line EFIT results.}
    \label{fig:kstarConfig}
\end{figure}

\reffig{fig:kstarConfig} shows locations where we obtain the input and the output data with the first wall (blue dotted line) on a poloidal cross-section of KSTAR. The green dotted line indicates a Rogowski coil measuring the plasma current ($I_\textrm{p}$).  The green open circles and crosses show locations of the magnetic pick-up coils measuring $32$ normal ($B_\textrm{n}$) and $36$ tangential ($B_\textrm{t}$) components of magnetic fields, respectively, whereas the green triangles show $22$ flux loops measuring the poloidal magnetic fluxes ($\Psi_\textrm{FL}$). These magnetic signals are selectively chosen out of all the magnetic sensors in KSTAR \cite{Lee:2008cl} whose performance has been demonstrated for many years, i.e., less susceptible to damages.

\begin{figure}
    \centering
    \includegraphics[width=1.0\linewidth]{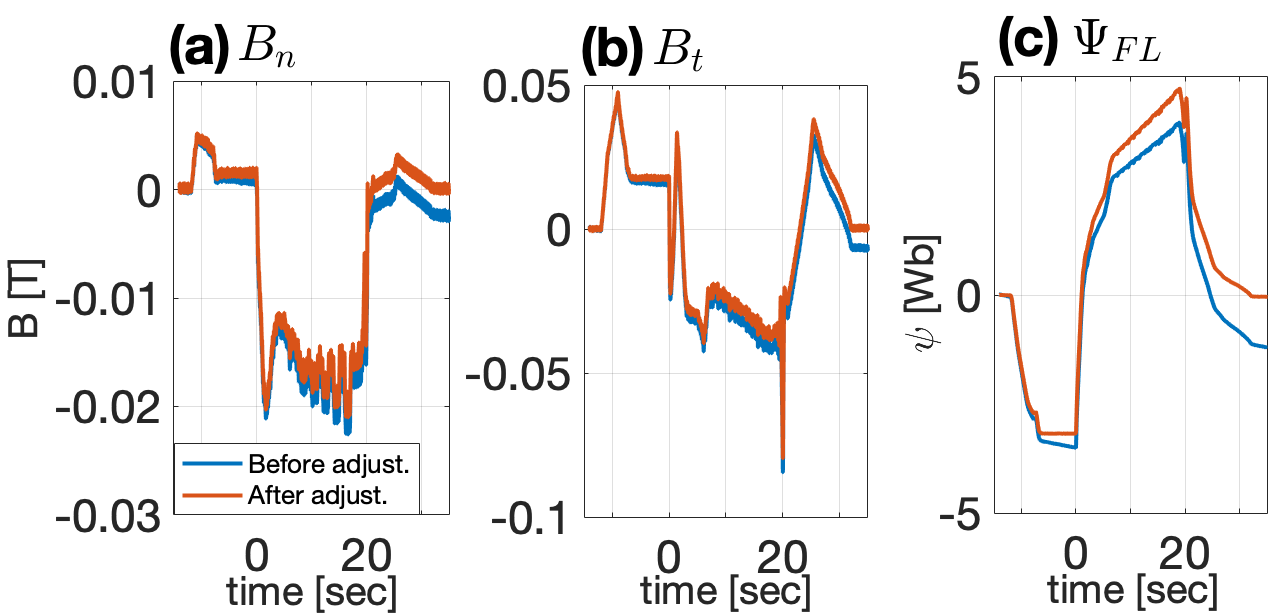}
    \caption{Before (blue) and after (red) the magnetic signal adjustments for (a) normal and (b) tangential components of magnetic fields measured by the magnetic pick-up coils, and (c) poloidal magnetic flux measured by one of the flux loops. The signals return closer to zeros after the adjustment when all the external magnetic coils (except the toroidal field coils) are turned off at around $30$ sec in this KSTAR discharge. See \refapp{app:drfit_adjust} for detailed description.}
    \label{fig:inputDrift}
\end{figure}

Although KSTAR calibrates the magnetic sensors (magnetic pick-up coils and flux loops) regularly during a campaign to remove drifts in the magnetic signals, it does not guarantee to fully eliminate such drifts. Thus, we preprocess the signals to adjust the drifts. \reffig{fig:inputDrift} shows examples of before (blue) and after (red) the drift adjustment for (a) normal and (b) tangential components of magnetic fields measured by the magnetic pick-up coils and (c) poloidal magnetic flux measured by one of the flux loops. Here, a KSTAR discharge is sustained until about $20$ sec, and all the external magnetic coils (except the toroidal field coils) are turned off at about $30$ sec. Therefore, we expect all the magnetic signals to return to zeros at around $30$ sec. If not, we envisage that there has been residual drifts. This means that we need to be able to preprocess the magnetic signals in real-time so that the input signal characteristics for predictions are similar to the trained ones. \refapp{app:drfit_adjust} describes in detail how we preprocess the magnetic signals in real-time.

The black asterisks in \reffig{fig:kstarConfig} show the $22\times 13$ grid points where we obtain the values of $\psi$ from the off-line EFIT results as outputs of the networks. We note that the original off-line EFIT provides the values of $\psi$ with $65\times 65$ grid points. The $22\times 13$ grid points are selected such that the distances between the neighboring channels in $R$ and $Z$ directions are as similar as possible while covering whole region within the first wall.  By generating such coarse grid points we can decrease the number of samples to train the network, thus consuming less amount of computational resources. Nevertheless, we do not lose the spatial resolution since $\lp R, Z\rp$ position is an input, i.e., the network can obtain the value of $\psi$ at any position within the first wall (see \refsec{S4:trainefit}).

\begin{table}
\centering
\caption{Summary of the data samples to train and validate the networks}
\label{tab:NNdata}
\resizebox{0.48\textwidth}{!}{%
\begin{tabular}{cccc}
\hline\hline
Parameter & Definition & \multicolumn{1}{l}{Data size} & \multicolumn{1}{l}{No. of samples} \\ \hline
$I_\textrm{p}$ & Plasma current & 1 &  \\
& (Rogowski coil) & & \\
 &  & \multicolumn{1}{l}{} & \multicolumn{1}{l}{} \\
$B_\textrm{n}$ & Normal magnetic field & 32 &  \\
& (Magnetic pick-up coils) & & \\
 &  &  & 217,820 \\
$B_\textrm{t}$ & Tangential magnetic field & 36 & (time slices) \\
& (Magnetic pick-up coils) & & \\
 &  &  &  \\
$\Psi_\textrm{FL}$ & Poloidal magnetic flux & 22 &  \\
& (Flux loops) & & \\
 &  &  &  \\ \hline
$R$ & Position in major radius & 1 & 286 \\
 &  &  & ($22\times13$ grids) \\
$Z$ & Position in height & 1 &  \\
 &  &  &  \\ \hline
Network Input size &  & 93 (+1 for bias) &  \\
 &  &  &  \\
Total no. of samples &  &  & 62,296,520 \\ \hline\hline
\end{tabular}%
}
\end{table}

With an additional input for the spatial position $R$ and $Z$, each data sample contains $93$ inputs (and yet another input for bias) and one output which is a value of $\psi$ at the specified $\lp R, Z\rp$ location. We randomly collect a total of $1,118$ KSTAR discharges from $2017$ and $2018$ campaigns. Since each discharge can be further broken into many time slices, i.e., every $50$ msec following the temporal resolution of the off-line EFIT, we obtain $217,820$ time slices. With a total of $286$ value of $\psi$ from $22\times 13$ spatial points, we have a total of $62,296,520\lp =217,820\times286\rp$ samples to train and validate the networks. $90$\% of the samples are used to train the networks, while the other $10$\% are used to validate the networks to avoid overfitting problems. Note that an overfitting problem can occur if a network is overly well trained to the \textit{training data} following the very details of them. This inhibits generalization of the trained network to predict \textit{unseen data}, and such a problem can be minimized with the validation data set. All the inputs except $R$ and $Z$ are normalized such that the maximum and minimum values within the whole samples become $1$ and $-1$, respectively. We use the actual values of $R$ and $Z$ in the unit of meters.

\reftable{tab:NNdata} summarizes the training and validation samples discussed in this section. Additionally, we also have randomly collected another $163$ KSTAR discharges in the same way discussed here which are different from the $1,118$ KSTAR discharges to test the performance of the networks.

\section{Neural network model and training} \label{S3:nn}

\subsection{Neural network model}

We develop the neural networks that not only output a value of $\psi$ but also satisfies \refeq{eq:gseq}, the GS equation. With the total of $94$ input nodes ($91$ for a plasma current and magnetic signals, two for $R$ and $Z$ position, one for the bias) and one output node for a value of $\psi$, each network has three fully connected hidden layers with an additional bias node at each hidden layer. Each layer contains $61$ nodes including the bias node. The structure of our networks is selected by examining several different structures by error and trials.

Denoting the value of $\psi$ calculated by the networks as $\psi^\textrm{NN}$, we have
\begin{equation}
\label{eq:nn_structure}
\begin{split}
&\psi^\textrm{NN}\! = \! s_{0} + \sum_{l=1}^{60}\! s_{l} \\
&\times f\!\! \left(\! u_{l0}\! + \sum_{k=1}^{60}\! u_{lk} f\!\! \left(\! v_{k0}\! +\! \sum_{j=1}^{60}\! v_{kj} f\!\! \left(\! w_{j0}\! +\! \sum_{i=1}^{93}\! w_{ji} x_{i}\! \right)\!\!\! \right)\!\!\! \right),
\end{split}
\end{equation}
where $x_i$ is the $i^\textrm{th}$ input value with $i=1,\dots,93$, i.e., $91$ measured values with the various magnetic diagnostics and two for $R$ and $Z$ positions. $w_{ji}$ is an element in a $61\times 94$ matrix, whereas $v_{kj}$ and $u_{lk}$ are elements in $61\times 61$ matrices. $s_l$ connects the $l^\textrm{th}$ node of the third (last) hidden layer to the output node. $w, v, u$ and $s$ are the weighting factors that need to be trained to achieve our goal of obtaining accurate $\psi$. $w_{j0}, v_{k0}, u_{l0}$ and $s_0$ are the weighting factors connecting the biases, where values of all the biases are fixed to be unity. We use a hyperbolic tangent function as the activation function $f$ giving the network non-linearity \cite{Haykin:2008}:
\begin{equation} \label{eq:NN-actfcn}
f\! \left(t \right) = \tanh(t) = \frac{2}{1 + e^{-2t}} - 1.
\end{equation}

The weighting factors are initialized as described in \cite{pmlr-v9-glorot10a} so that a good training can be achieved. They are randomly selected from a normal distribution whose mean is zero with the variance set to be an inverse of total number of connecting nodes. For instance, our weighting factor $w$ connects the input layer (94 nodes with bias) and the first hidden layer (61 nodes with bias), therefore the variance is set to be $1/(94+61)$. Likewise, the variances for $v$, $u$ and $s$ are $1/(61+61)$, $1/(61+61)$ and $1/(61+1)$, respectively.

\subsection{Training}

With the aforementioned network structure, training (or optimizing) the weighting factors to predict the correct value of $\psi$ highly depends on a choice of the cost function. A typical choice of such cost function would be:
\begin{equation} \label{eq:costfcn}
\epsilon = \frac{1}{N} \sum_{i=1}^{N} \left( \psi_{i}^\textrm{NN} - \psi_{i}^\textrm{Target} \right)^{2},
\end{equation}
where $\psi^\textrm{Target}$ is the target value, i.e., the value of $\psi$ from the off-line EFIT results in our case, and $N$ the number of data sets.

As will be shown shortly, minimizing the cost function $\epsilon$ does not guarantee to satisfy the GS equation (\refeq{eq:gseq}) even if $\psi^\textrm{NN}$ and $\psi^\textrm{Target}$ matches well, i.e., the network is well trained with the given optimization rule. Since $\Delta^*\psi$ provides information on the toroidal current density directly, it is important that $\Delta^*\psi^\textrm{NN}$ matches $\Delta^*\psi^\textrm{Target}$ as well. We have an analytic form representing $\psi^\textrm{NN}$ as in \refeq{eq:nn_structure}; therefore, we can analytically differentiate $\psi^\textrm{NN}$ with respect to $R$ and $Z$, meaning that we can calculate $\Delta^*\psi^\textrm{NN}$ during the training stage. Thus, we introduce another cost function:
\begin{equation} \label{eq:costfcn_plusdiff}
\begin{split}
\epsilon^\textrm{new}& = \frac{1}{N} \sum_{i=1}^{N} \left( \psi_{i}^\textrm{NN} - \psi_{i}^\textrm{Target} \right)^{2} \\
&+ \frac{1}{N} \sum_{i=1}^{N} \left( \Delta^*\psi_{i}^\textrm{NN} - \Delta^*\psi_{i}^\textrm{Target} \right)^{2},
\end{split}
\end{equation}
where we obtain the value of $\Delta^*\psi^\textrm{Target}$ from the off-line EFIT results as well. 

\begin{figure}
    \centering
    \includegraphics[width=0.95\linewidth]{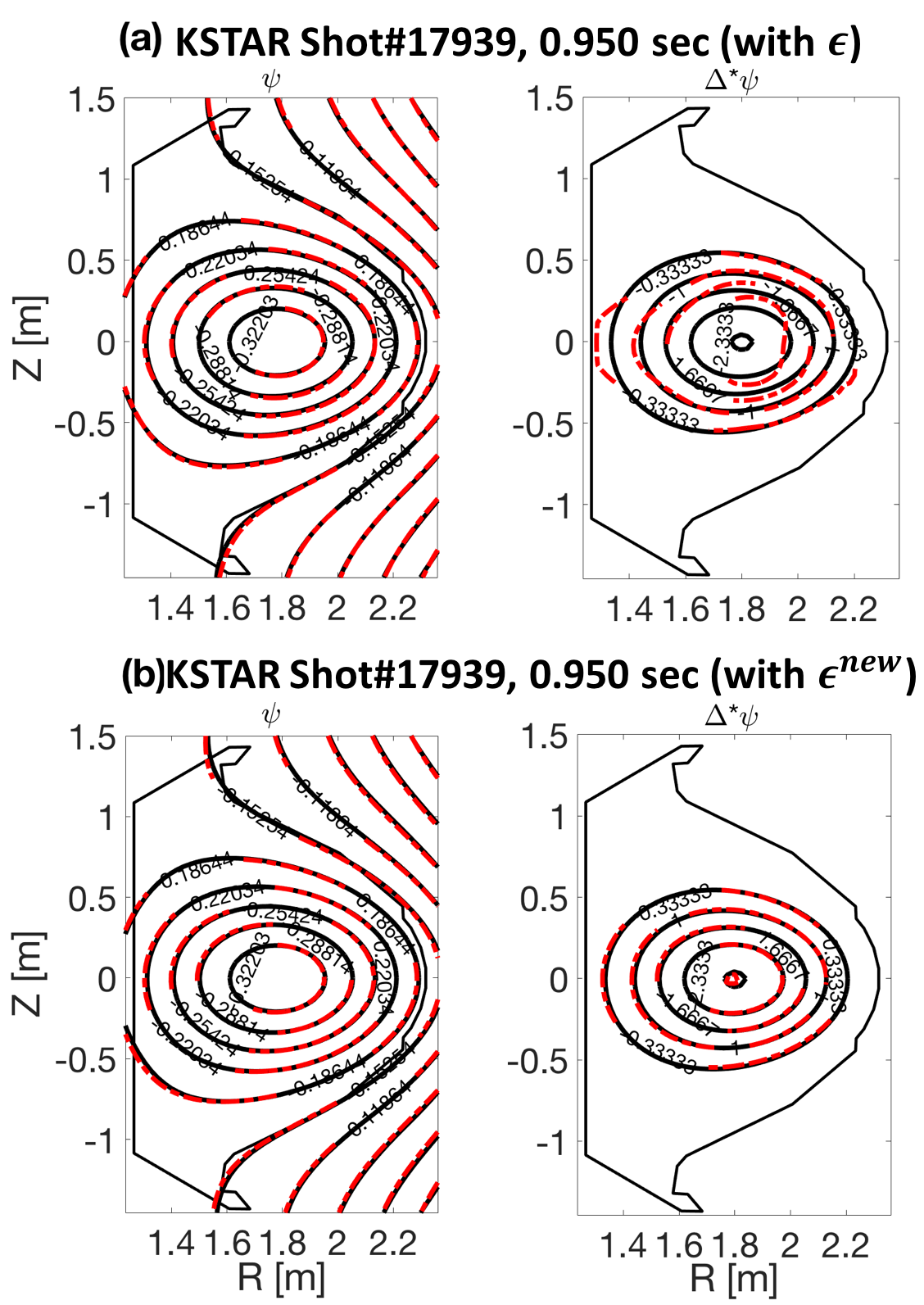}
    \caption{An example of the two networks' results trained with the cost function (a) $\epsilon$ and (b) $\epsilon^\textrm{new}$ for KSTAR shot\# $17939$ at $0.950$ sec. Both networks (red dashed line) reproduce the $\psi^\textrm{Target}$ (black line) well (left panels), but only the network trained with $\epsilon^\textrm{new}$ reproduces $\Delta^*\psi^\textrm{Target}$ (right panels).}
    \label{fig:costco}
\end{figure}

To acknowledge difference between the two cost functions $\epsilon$ and $\epsilon^\textrm{new}$, we first discuss the results. \reffig{fig:costco} shows the outputs of the two trained networks with the cost function (a) $\epsilon$ and (b) $\epsilon^\textrm{new}$. It is evident that in both cases the network output $\psi^\textrm{NN}$ (red dashed line) reproduces the off-line EFIT $\psi^\textrm{Target}$ (black line). However, only the network trained with the cost function $\epsilon^\textrm{new}$ reproduces the off-line EFIT $\Delta^*\psi^\textrm{Target}$. Both networks are trained well, but the network with the cost function $\epsilon$ does not achieve our goal, that is correctly predicting $\psi^\textrm{Target}$ \textit{and} $\Delta^*\psi^\textrm{Target}$.

Since our goal is to develop a neural network that solves the GS equation, we choose the cost function to be $\epsilon^\textrm{new}$ to train the networks. We optimize the weighting factors by minimizing $\epsilon^\textrm{new}$ with the Adam \cite{DBLP:journals/corr/KingmaB14} which is one of the gradient-based optimization algorithms. With $90$\% and $10$\% of the total data samples for training and validation of the networks, respectively, we stop training the networks with a fixed number of iterations that is large enough but not too large such that the validation errors do not increase, i.e., to avoid overfitting problems. The whole workflow is carried out with Python and Tensorflow \cite{tensorflow2015-whitepaper}.

With the selected cost function we create three different networks that differ only by the training data sets. NN$_\textrm{2017}$, NN$_\textrm{2018}$ and NN$_\textrm{2017, 2018}$ refer to the three networks trained with the data sets from only $2017$ ($744$ discharges), from only $2018$ ($374$ discharges) and from both $2017$ and $2018$ ($744+374$ discharges) campaigns, respectively. 

\begin{figure}
    \centering
    \includegraphics[width=0.95\linewidth]{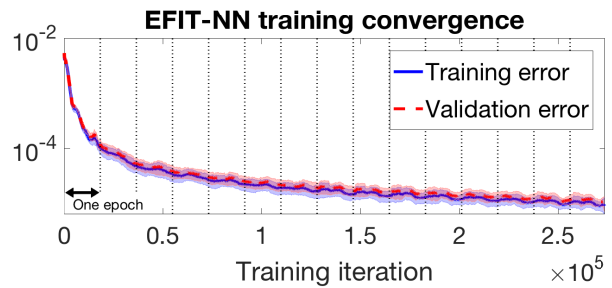}
    \caption{The descending feature of training (blue line) and validation (red dashed line) errors as a function of iterations. Shaded areas represent standard deviation of the errors.}
    \label{fig:conv}
\end{figure}

The descending feature of the cost function $\epsilon^\textrm{new}$ as a function of the training iteration for NN$_{2017, 2018}$ network is shown in \reffig{fig:conv}. Both the training errors (blue line) and validation errors (red dashed line) decrease together with similar values which means that the network is well generalized. Furthermore, since the validation errors do not increase, the network does not have an overfitting problem. Note that fluctuations in the errors, i.e., standard deviation of the errors, are represented as shaded areas. 

Small undulations repeated over the iterations in \reffig{fig:conv} are due to the mini-batch learning. Contrary to the batch learning, i.e., optimizing the network with the entire training set in one iteration, the mini-batch learning divides the training set into some number of small subsets ($1,000$ subsets for our case) to optimize the networks sequentially. One cycle that goes through all the subsets once is called an epoch. The mini-batch learning helps to escape from local minima in the weighting factor space \cite{DBLP:journals/corr/GeHJY15} via the stochastic gradient descent scheme \cite{Bottou10large-scalemachine}.

\section{Performance of the developed neural networks: Benchmark tests} \label{S4:trainefit}

In this section, we present how well the developed networks perform. Main figures of merit we use are peak signal-to-noise ratio (PSNR) and mean structural similarity (MSSIM) as have been used perviously \cite{Matos:2017kl} in addition to the usual statistical quantity R$^2$, coefficient of determination. We note that obtaining full flux surface information $\psi\lp R, Z\rp$ on $22\times 13$ or $65\times 65$ spatial grids with our networks takes less than $1$ msec on a typical personal computer. 

First, we discuss the benchmark results of the NN$_{2017, 2018}$ network. Then, we compare the performance of NN$_{2017}$, NN$_{2018}$ and NN$_{2017, 2018}$ networks. Here, we also investigate cross-year performance, for instance, applying the NN$_{2017}$ network to predict the discharges obtained from 2018 campaign and vice versa. Then, we evaluate the performance of the networks against the rt-EFIT results to examine possibility of supplementing or even replacing the rt-EFIT with the networks. Finally, we show how the imputation scheme supports the networks' performance. Here, all the tests are performed with the unseen (to all three networks, i.e., NN$_{2017}$, NN$_{2018}$ and NN$_{2017, 2018}$) KSTAR discharges which are $88$ and $75$ KSTAR discharges from 2017 and 2018 campaigns, respectively.

\subsection{Benchmark results of the NN$_{2017, 2018}$ network} \label{S4-1:benchmark}

\begin{figure}
    \centering
    \includegraphics[width=0.95\linewidth]{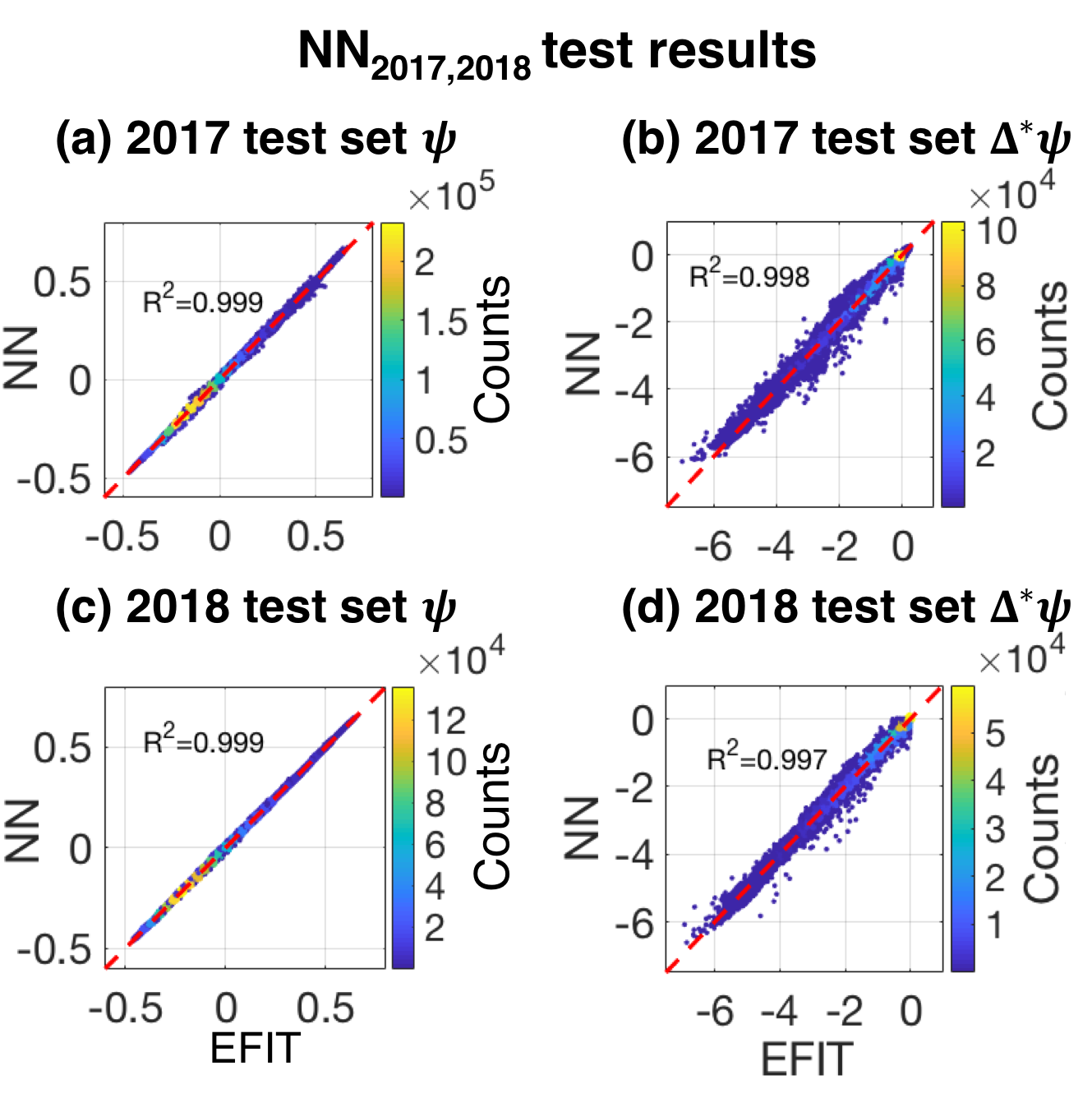}
    \caption{Performance tests of the NN$_{2017, 2018}$ network on the unseen KSTAR discharges from (a)(b) 2017 campaign and (c)(d) 2018 campaign. The values of R$^2$ and histograms of (a)(c) $\psi^\textrm{NN}$ vs. $\psi^\textrm{Target}$ and (b)(d) $\Delta^*\psi^\textrm{NN}$ vs. $\Delta^*\psi^\textrm{Target}$ with colors representing number of counts manifest goodness of the NN$_{2017, 2018}$ network. Red dashed line is the $y=x$ line.}
    \label{fig:teLinear1718}
\end{figure}

\begin{figure*}
    \centering
    \includegraphics[width=0.95\linewidth]{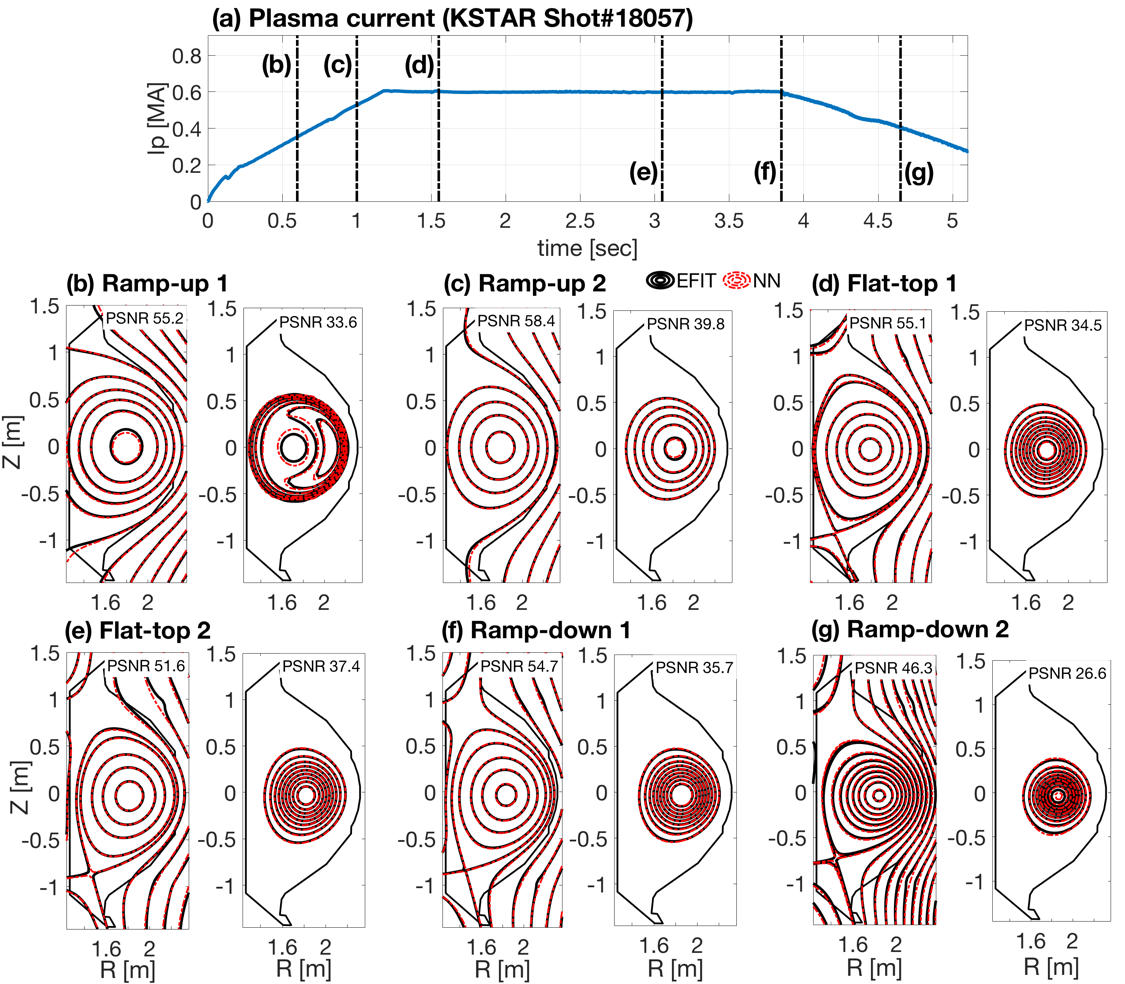}
    \caption{The actual reconstruction results for the KSTAR shot\#18057, comparing the network results and off-line EFIT reconstructions for ramp-up ((b) and (c)), flat-top ((d) and (e)), ramp-down ((f) and (g)) phases following (a) the plasma current evolution. Black lines indicate the flux surfaces from the off-line EFIT, overlaid with the red dotted lines which stand for the NN reconstructions. As a figure of merit, magnitudes of PSNR metric are written on each figure.}
    \label{fig:teFSshot}
\end{figure*}

\reffig{fig:teLinear1718} show the benchmark results of the NN$_{2017, 2018}$ network, i.e., network trained with the data sets from both 2017 and 2018 campaigns. (a) and (b) show the results with the test discharges from 2017 campaign; while (c) and (d) present the results with the test discharges from 2018 campaign. Histograms of (a)(c) $\psi^\textrm{NN}$ vs. $\psi^\textrm{Target}$ and (b)(d) $\Delta^*\psi^\textrm{NN}$ vs. $\Delta^*\psi^\textrm{Target}$ are shown with colors representing the number of counts. \correction{For instance, there is a yellow colored point in \reffig{fig:teLinear1718}(a) around $(-0.1, -0.1)\pm\varepsilon$, where $\varepsilon$ is a bin size for the histogram. Since yellow represents about $2\times 10^5$ counts, there are approximately $2\times 10^5$ data whose neural network values and EFIT values are $-0.1\pm\varepsilon$ simultaneously within our test data set.} Note that each KSTAR discharge contains numerous time slices whose number depends on the actual pulse length of a discharge, and each time slice generates the total of $22\times 13=286$ data points. The values of $\psi^\textrm{Target}$ and $\Delta^*\psi^\textrm{Target}$ are obtained from the off-line EFIT results. It is clear that the network predicts the target values well. 

As a figure of merit, we introduce the R$^2$ metric (coefficient of determination) defined as 
\begin{equation}
\textrm{R}^2 = 1 - \frac{\sum_{i=1}^{L} \left(y_i^\textrm{Target} - y_i^\textrm{NN} \right)^2}{\sum_{i=1}^{L} \left(y_i^\textrm{Target} - \frac{1}{L}\sum_{j=1}^{L} y_{j}^\textrm{Target} \right)^2},
\end{equation}
where $y$ takes either $\psi$ or $\Delta^*\psi$, and $L$ is the number of test data sets. The calculated values are written in \reffig{fig:teLinear1718}, and they are indeed close to unity, implying the existence of very strong linear correlations between the predicted (from the network) and target (from the off-line EFIT) values. Note that R$^2=1$ means the perfect prediction. The red dashed lines on the figures are the $y=x$ lines.

\reffig{fig:teFSshot} is an example of reconstructed magnetic equilibria using KSTAR shot \#18057 from 2017 campaign. (a) shows the evolution of the plasma current. The vertical dashed lines indicate the time points where we show and compare the equilibria obtained from the network (red) and the off-line EFIT (black) which is our target. (b) and (c) are taken during the ramp-up phase, (d) and (e) during the flat-top phase, and (f) and (g) during the ramp-down phase. In each sub-figure from (b) to (g), left panels compare $\psi$, and right panels are for $\Delta^*\psi$. We mention that the equilibria in \reffig{fig:teFSshot} are reconstructed with $65 \times 65$ grid points even though the network is trained with $22 \times 13$ grid points demonstrating how spatial resolution is flexible in our networks.

For a quantitative assessment of the network, we use an image relevant figure of merit that is peak signal-to-noise ratio (PSNR) \cite{HuynhThu:2008fm} (see \refapp{app:image_figures}) originally developed to estimate a degree of artifacts due to an image compression compared to an original image. Typical PSNR range for the JPEG image, which preserves the original quality with a reasonable degree, is generally in 30--50 dB \cite{Matos:2017kl, Ebrahimi:2004fz}. For our case, the networks errors relative to the off-line EFIT results can be treated as artifacts. As listed on \reffig{fig:teFSshot}(b)-(g), PSNR for $\psi$ is very good, while we achieve acceptable values for $\Delta^*\psi$.

\subsection{The NN$_{2017}$, NN$_{2018}$ and NN$_{2017, 2018}$ networks}

\begin{figure}
    \centering
    \includegraphics[width=0.95\linewidth]{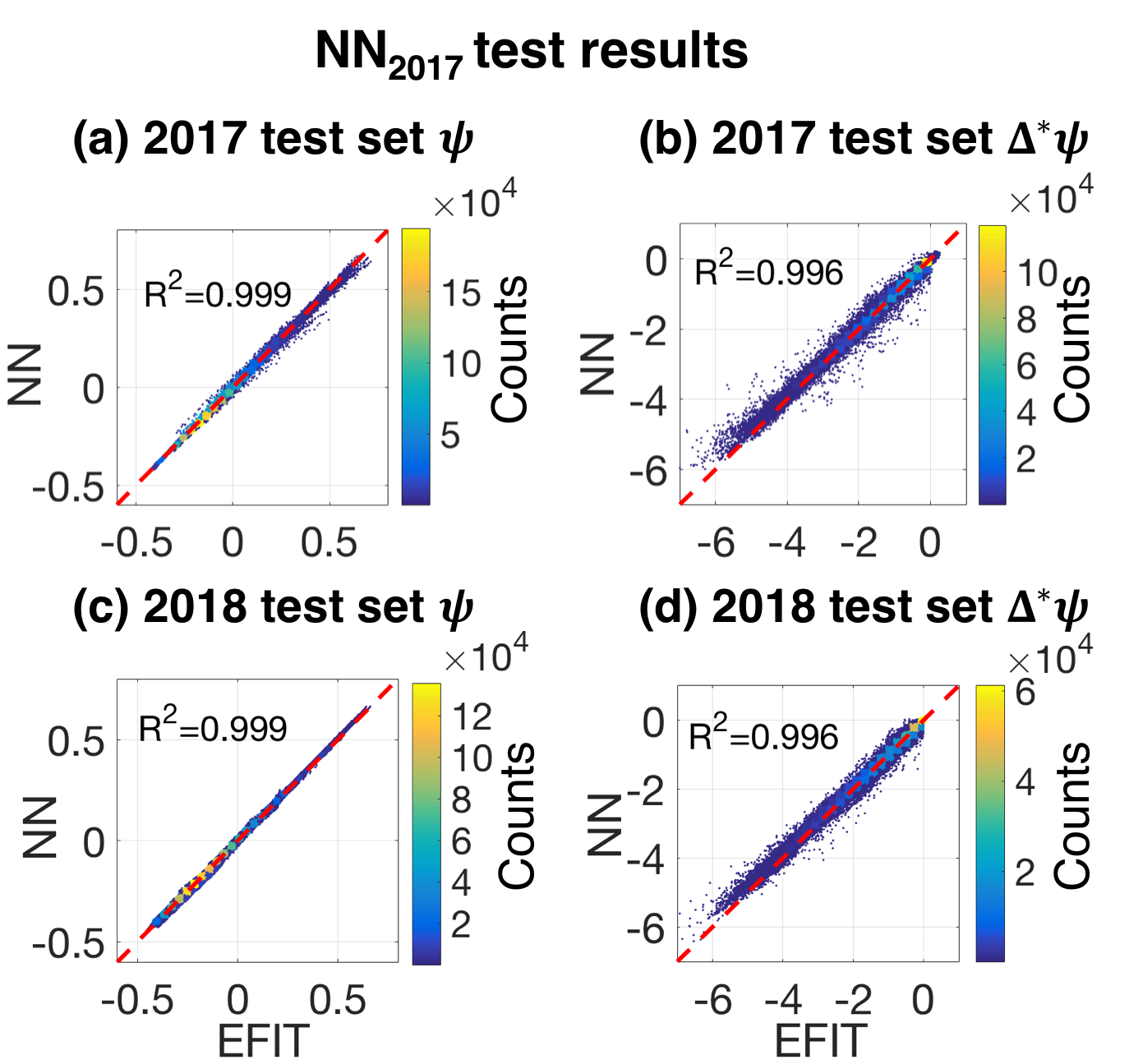}
    \caption{Same as \reffig{fig:teLinear1718} for the the NN$_{2017}$ network, i.e., trained with the data sets from 2017 campaign.}
    \label{fig:teLinear17}
\end{figure}

\begin{figure}
    \centering
    \includegraphics[width=0.95\linewidth]{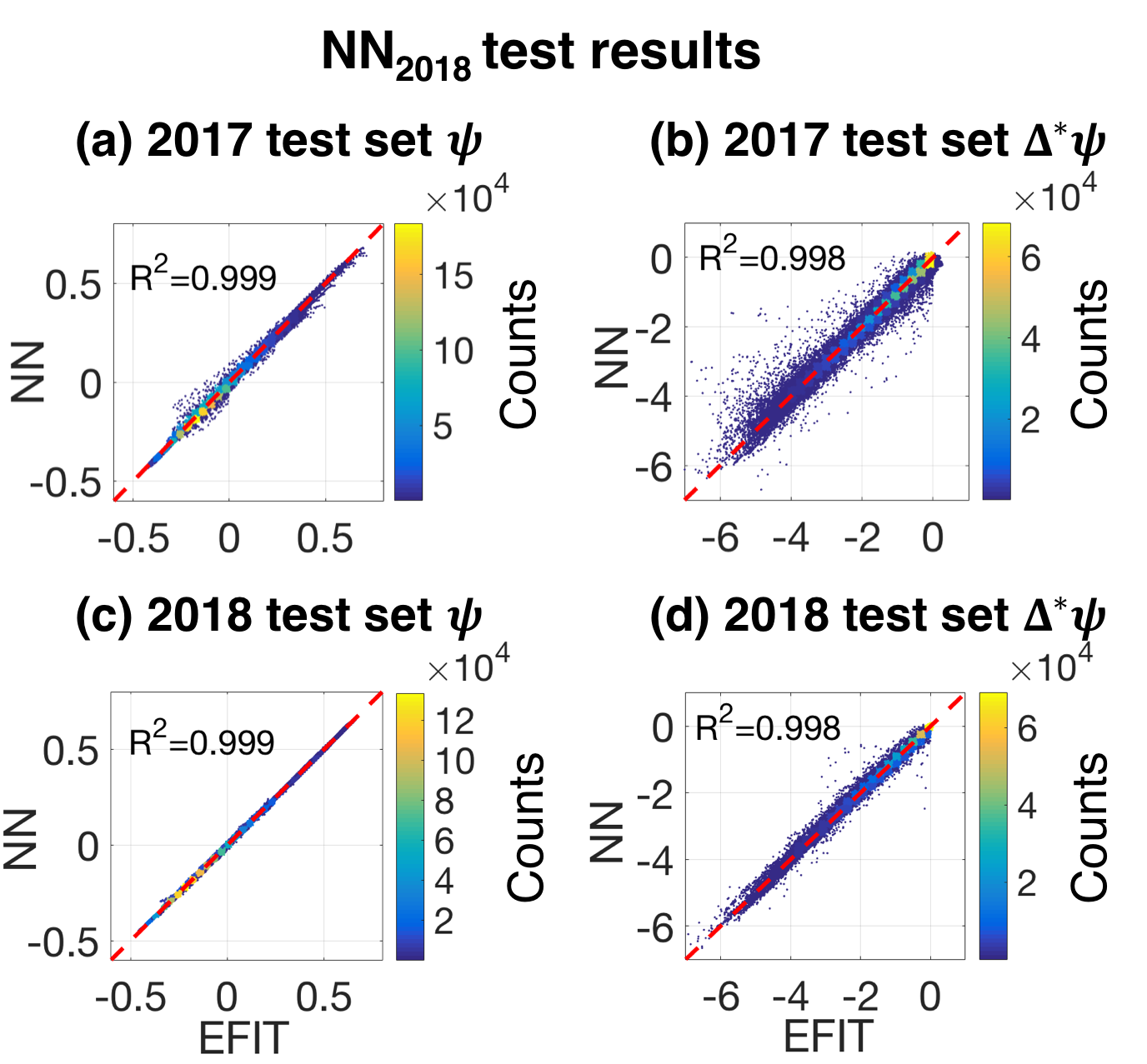}
    \caption{Same as \reffig{fig:teLinear1718} for the the NN$_{2018}$ network, i.e., trained with the data sets from 2018 campaign.}
    \label{fig:teLinear18}
\end{figure}

Similar to shown in \reffig{fig:teLinear1718}, we show the benchmark results of the NN$_{2017}$ (trained with the data sets from 2017 campaign) and the NN$_{2018}$ (trained with the data sets from 2018 campaign) in \refmultifig{fig:teLinear17}{fig:teLinear18}, respectively. R$^2$ metric is also provided on the figures. Again, overall performance of the networks are good. 

The NN$_{2017}$ and NN$_{2018}$ networks are trained with only in-campaign data sets, e.g., NN$_{2018}$ with the data sets from only 2018 campaign, and we find slightly worse results, but still good, on predicting cross-campaign magnetic equilibria, e.g. NN$_{2018}$ predicting equilibria for 2017 campaign. Notice that the NN$_{2017}$ seems to predict cross-campaign equilibria better than in-campaign ones by comparing \reffig{fig:teLinear17}(a) and (c) which contradicts our intuition. Although the histogram in \reffig{fig:teLinear17}(c) seems tightly aligned with the $y=x$ line (red dashed line), close inspection reveals that the NN$_{2017}$ network, in general, underestimates the off-line EFIT results from 2018 campaign marginally. This will be evident when we compare image qualities.

Mean structural similarity (MSSIM) \cite{Wang:2004gj} (see \refapp{app:image_figures}) is another image relevant figure of merit used to estimate perceptual  similarity (or perceived differences) between the true and reproduced images based on inter-dependence of adjacent spatial pixels in the images. MSSIM ranges from zero to one, where the closer to unity the better the reproduced image is.

\begin{figure}
    \centering
    \includegraphics[width=1\linewidth]{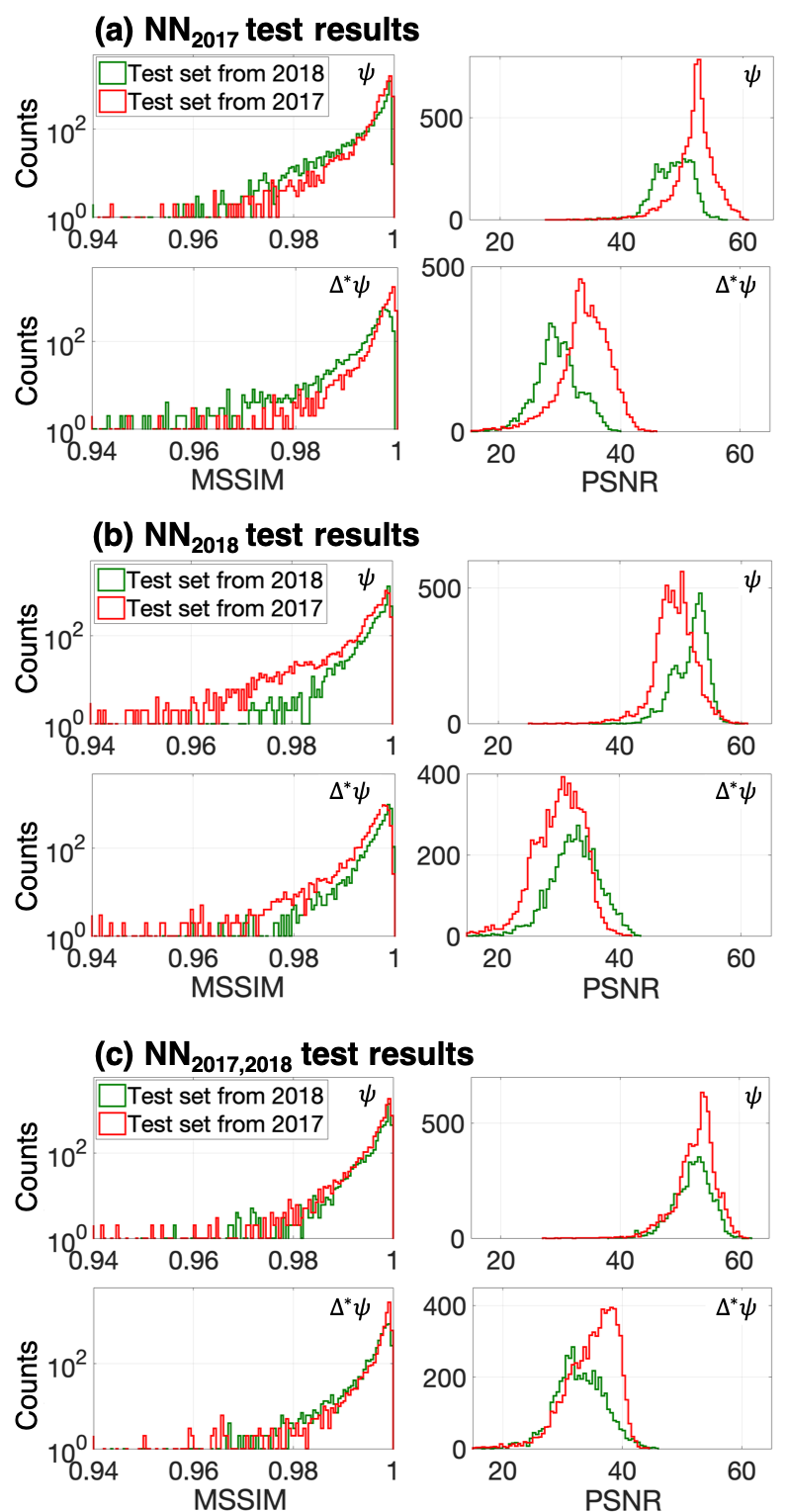}
    \caption{Histograms of MSSIM (left panel) and PSNR (right panel) for (a) NN$_{2017}$, (b) NN$_{2018}$ and (c) NN$_{2017, 2018}$. Red (green) line indicates the test results on the data sets from 2017 (2018) campaign. In each sub-figure, top (bottom) panel show the results for $\psi$ ($\Delta^*\psi$). The off-line EFIT results are used as reference.}
    \label{fig:tedist}
\end{figure}

Together with PSNR, \reffig{fig:tedist} shows MSSIM for (a) NN$_{2017}$, (b) NN$_{2018}$ and (c) NN$_{2017, 2018}$ where the off-line EFIT results are used as reference. Notice that counts in all the histograms of MSSIM and PSNR in this work correspond to the number of reconstructed magnetic equilibria (or a number of time slices) since we obtain a single value of MSSIM and PSNR from one equilibrium; whereas counts in Figures \ref{fig:teLinear1718}, \ref{fig:teLinear17} and \ref{fig:teLinear18} are much bigger since $286(=22\times 13)$ data points are generated from each time slice. Red (green) line indicates the test results on the data sets from 2017 (2018) campaign. In general, whether the test data sets are in-campaign or cross-campaign, image reproducibility of all three networks, i.e., predicting the off-line EFIT results, is good as attested by the fact that MSSIM is quite close to unity and PSNR for $\psi$ ($\Delta^*\psi$) ranges approximately $40$ to $60$ ($20$ to $40$). It is easily discernible that in-campaign results are better for both NN$_{2017}$ and NN$_{2018}$ unlike what we noted in \reffig{fig:teLinear17}(a) and (c). Not necessarily guaranteed, we find that the NN$_{2017, 2018}$ network works equally well for both campaigns as shown in \reffig{fig:tedist}(c).

\subsection{Comparisons among nn-EFIT, rt-EFIT and off-line EFIT} \label{S4-2:rtefit}

\begin{figure}
    \centering
    \includegraphics[width=0.70\linewidth]{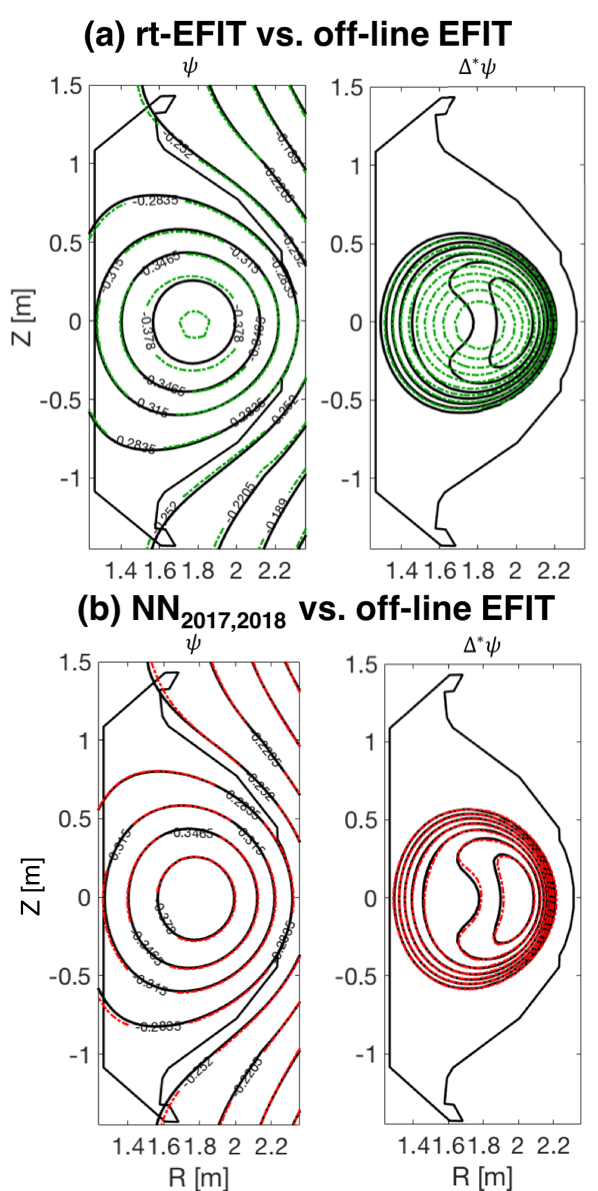}
    \caption{An example of reconstructed $\psi\lp R, Z\rp$ (left panel) and $\Delta^*\psi\lp R, Z\rp$ (right panel) for KSTAR shot \#$17975$ at $0.7$ sec comparing (a) rt-EFIT (green) and off-line EFIT (black) and (b) nn-EFIT (NN$_{2017, 2018}$) (red) and off-line EFIT (black).}
    \label{fig:rtFSshot}
\end{figure}

It is widely recognized that rt-EFIT results and off-line results are different from each other. If we allow the off-line EFIT results used to train the networks to be accurate ones, then the reconstruction of equilibria with the neural networks (nn-EFIT) must satisfy the following criterion: nn-EFIT results must be more similar to the off-line EFIT results than rt-EFIT results are to the off-line EFIT as mentioned in \refsec{S1:intro}. Once this criterion is satisfied, then we can always improve the nn-EFIT as genuinely more accurate EFIT results are collected. For this reason, we make comparisons among the nn-EFIT, rt-EFIT and off-line EFIT results. 

\reffig{fig:rtFSshot} shows an example of reconstructed magnetic equilibria for (a) rt-EFIT vs. off-line EFIT and (b) nn-EFIT (the NN$_{2017, 2018}$ network) vs. off-line EFIT for KSTAR shot \#$17975$ at $0.7$ sec with $\psi$ (left panel) and $\Delta^*\psi$ (right panel). Green, red and black lines indicate rt-EFIT, nn-EFIT and off-line EFIT results, respectively. This simple example shows that the nn-EFIT is more similar to the off-line EFIT than the rt-EFIT is to the off-line EFIT, satisfying the aforementioned criterion.

\begin{figure}
    \centering
    \includegraphics[width=0.9\linewidth]{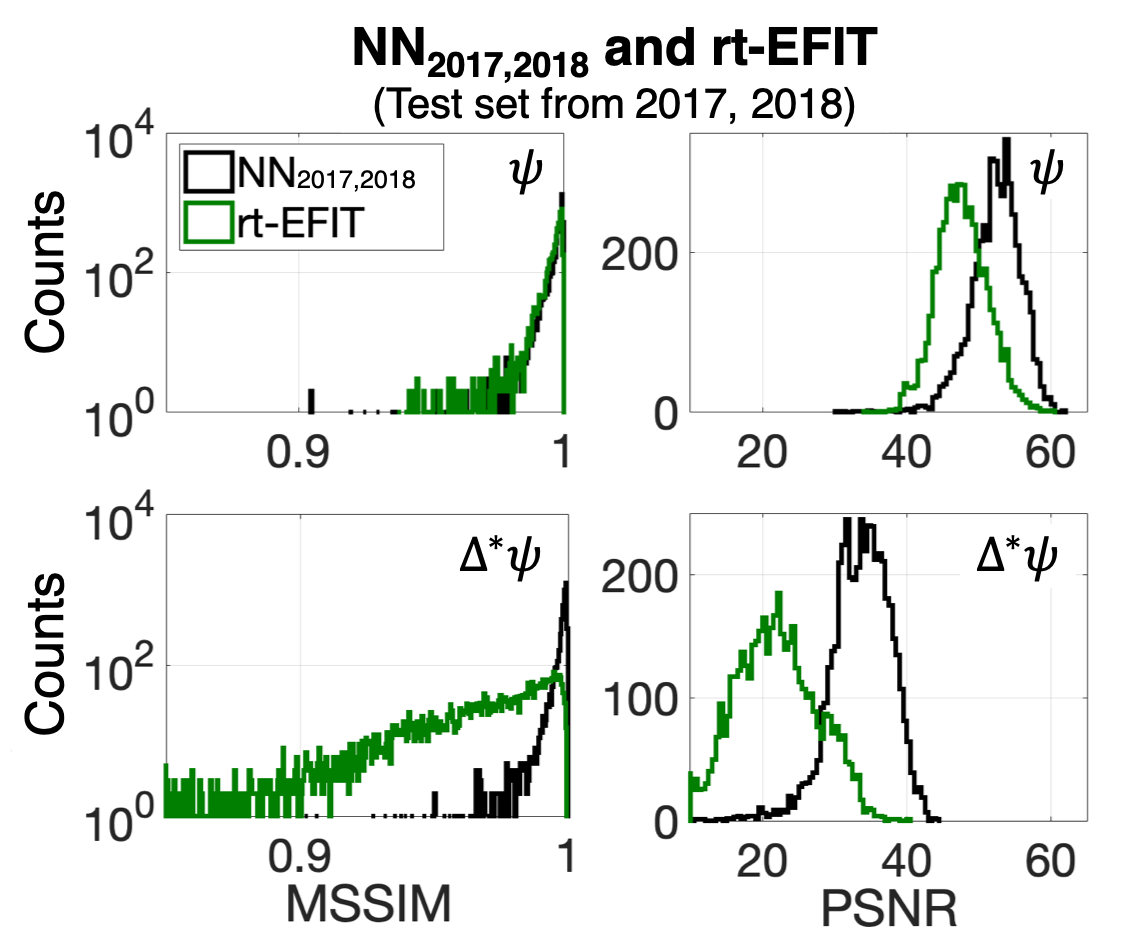}
    \caption{Histograms of MSSIM (left panel) and PSNR (right panel) of $\psi$ (top) and $\Delta^*\psi$ (bottom) calculated by the nn-EFIT (black) and the rt-EFIT (green), where the nn-EFIT is the NN$_{2017, 2018}$. For both the nn-EFIT and the rt-EFIT, the off-line EFIT is treated as reference.}
    \label{fig:rtEFIT_nn_2017_2018}
\end{figure}

To validate the criterion statistically, we generate histograms of MSSIM and PSNR for the nn-EFIT and the rt-EFIT with reference to the off-line EFIT. This is shown in \reffig{fig:rtEFIT_nn_2017_2018} as histograms, where MSSIM (left panel) and PSNR (right panel) of $\psi$ (top) and $\Delta^*\psi$ (bottom) are compared between the nn-EFIT (black) and the rt-EFIT (green). Here, the nn-EFIT results are obtained with the NN$_{2017, 2018}$ network on the test data sets. We confirm that the criterion is satisfied with the NN$_{2017, 2018}$ network as the histograms in \reffig{fig:rtEFIT_nn_2017_2018} are in favour of the nn-EFIT, i.e., larger MSSIM and PSNR are obtained by the nn-EFIT. This is more conspicuous for $\Delta^*\psi$ than $\psi$.

\begin{figure}
    \centering
    \includegraphics[width=0.9\linewidth]{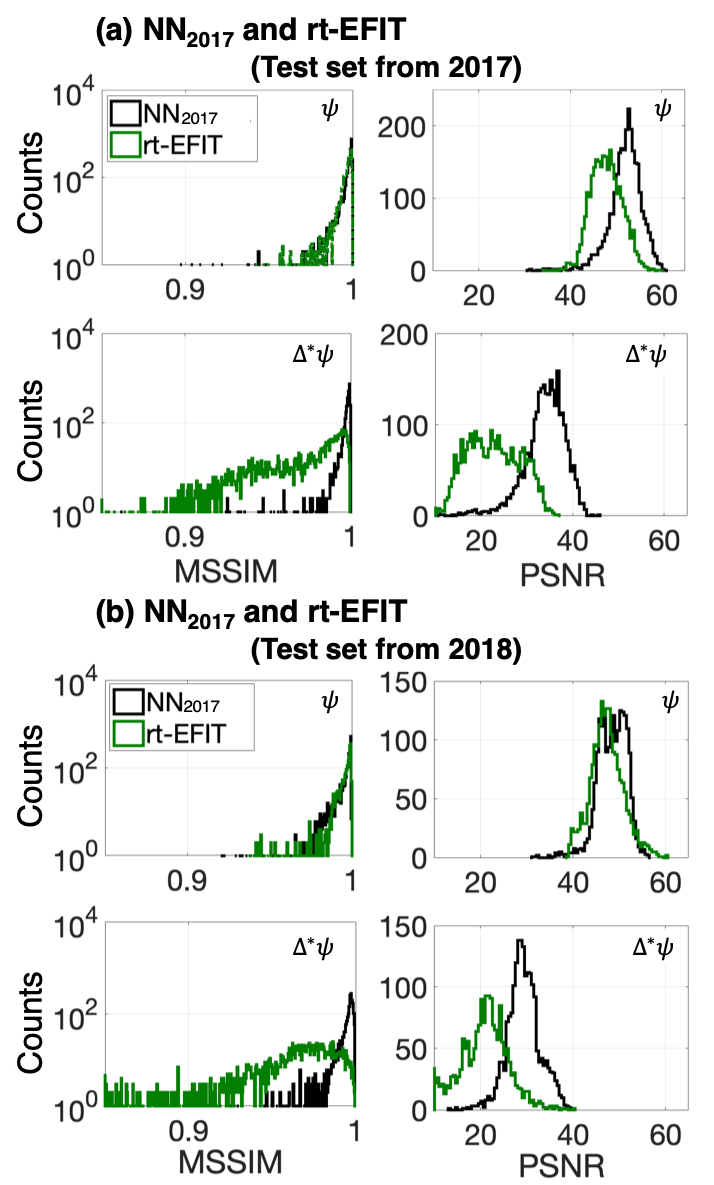}
    \caption{Same as \reffig{fig:rtEFIT_nn_2017_2018} with the NN$_{2017}$ as the nn-EFIT where the test data sets are obtained from (a) 2017 campaign and (b) 2018 campaign.}
    \label{fig:rtEFIT_nn_2017}
\end{figure}

\begin{figure}
    \centering
    \includegraphics[width=0.9\linewidth]{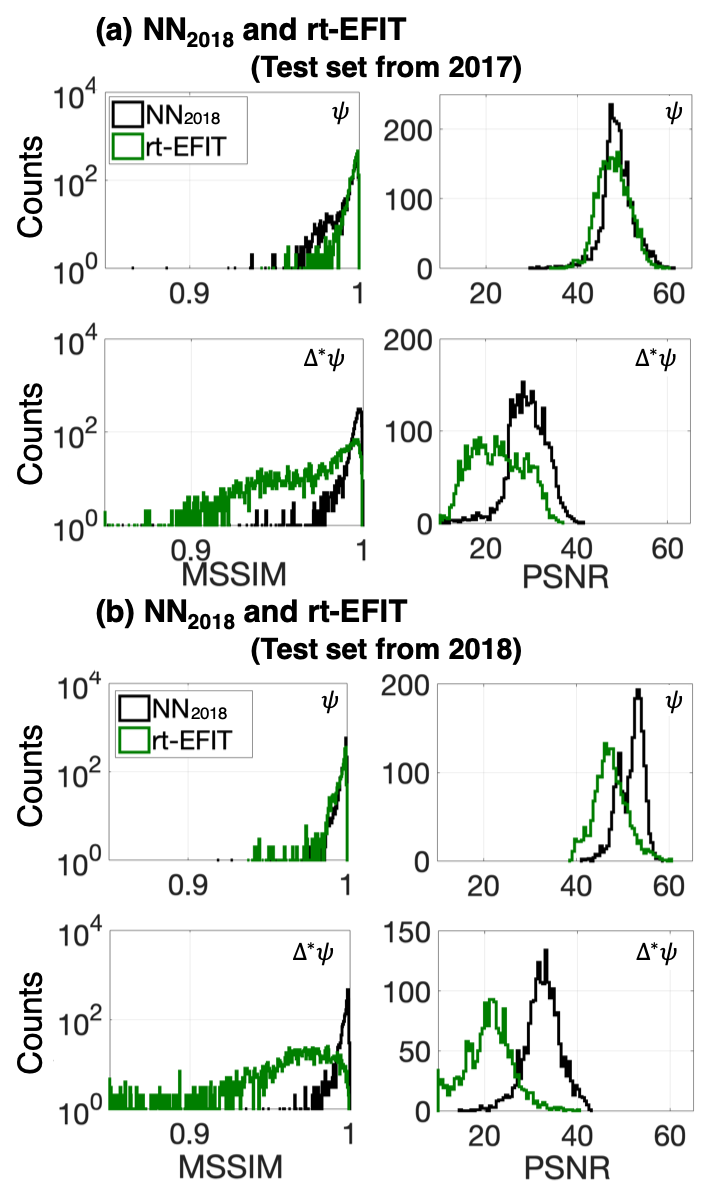}
    \caption{Same as \reffig{fig:rtEFIT_nn_2017_2018} with the NN$_{2018}$ as the nn-EFIT where the test data sets are obtained from (a) 2017 campaign and (b) 2018 campaign.}
    \label{fig:rtEFIT_nn_2018}
\end{figure}

We perform the similar statistical analyses for the other two networks, NN$_{2017}$ and NN$_{2018}$, which are shown in \refmultifig{fig:rtEFIT_nn_2017}{fig:rtEFIT_nn_2018}. Since these two networks are trained with the data sets from only one campaign, we show the results where the test data sets are prepared from (a) 2017 campaign or (b) 2018 campaign so that in-campaign and cross-campaign effects can be assessed separately. We find that whether in- or cross-campaign, the criterion is fulfilled for both $\psi$ and $\Delta^*\psi$.

\subsection{The NN$_{2017, 2018}$ network with the imputation scheme} \label{S5:imputation}

If one or a few magnetic pick-up coils which are a part of the inputs to the nn-EFIT are impaired, then we will have to re-train the network without the damaged ones or hope that the network will reconstruct equilibria correctly by padding a fixed value, e.g., zero-padding, to the broken ones. Of course, one can anticipate training the network by considering possible combinations of impaired magnetic pick-up coils. With the total number of $68$ signals from the magnetic pick-up coils being inputs to the network in our case, we immediately find that the number of possible combinations increases too quickly to consider it as a solution. 

Since inferring the missing values is better than the null replacement \cite{vanLint:2005cn}, we resolve the issue by using the recently proposed imputation method \cite{Joung:2018ju} based on Gaussian processes (GP) \cite{Rasmussen:2006} and Bayesian inference \cite{Sivia:2006}, where the likelihood is constructed based on Maxwell's equations. The imputation method infers the missing values fast enough, i.e., less than $1$ msec to infer at least up to nine missing values on a typical personal computer; thus, we can apply the method during a plasma discharge by replacing the missing values with the real-time inferred values. 

\begin{figure}
    \centering
    \includegraphics[width=0.8\linewidth]{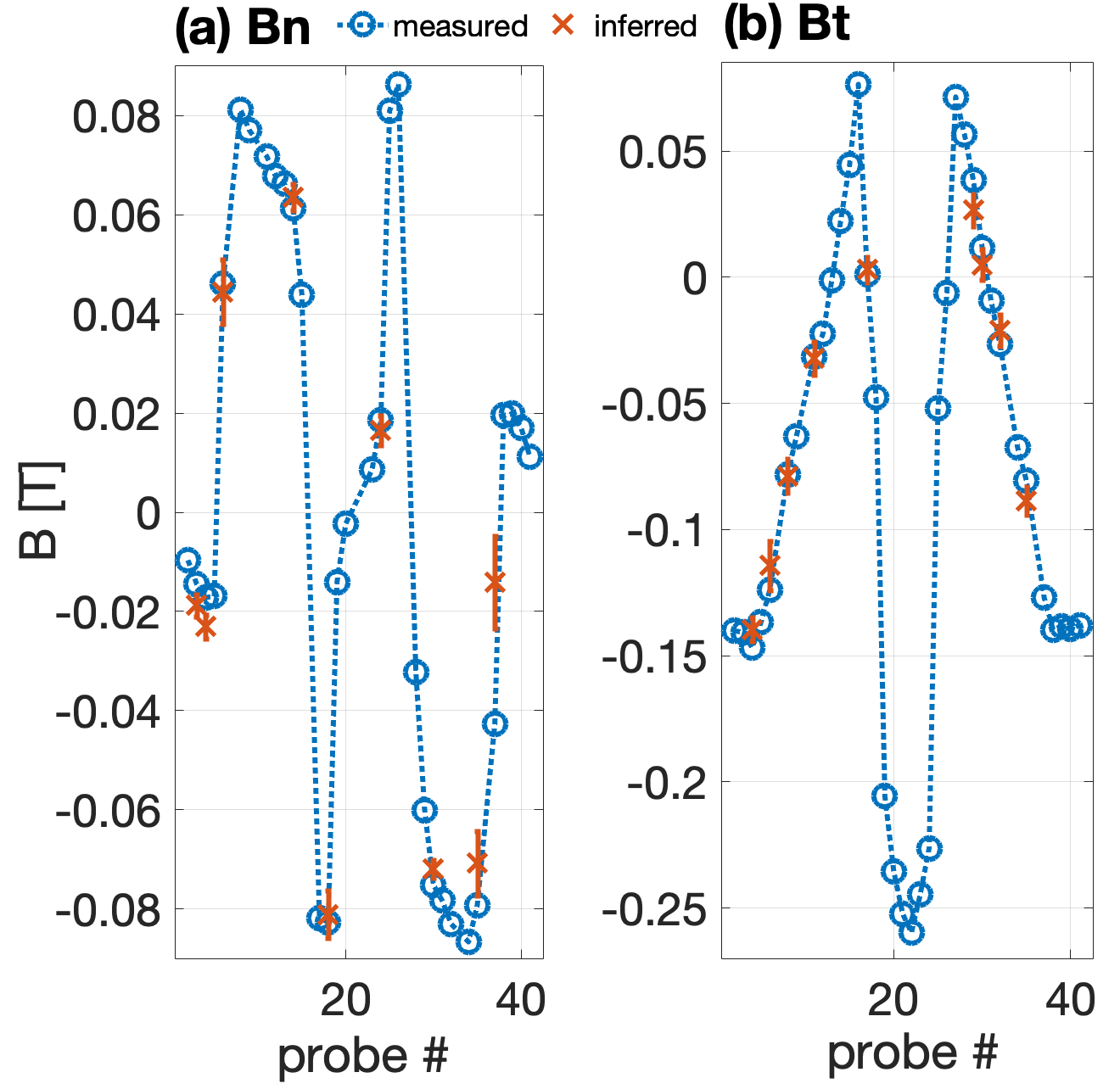}
    \caption{Measured (blue open circles) and inferred with the imputation method \cite{Joung:2018ju} (red crosses with their uncertainties) values for (a) $B_\textrm{n}$ and (b) $B_\textrm{t}$. Probe \# on the horizontal axis is used as an identification index of magnetic pick-up coils. Inferred probes are Probe \#3, 4, 6, 14, 18, 24, 30, 35, 37 for $B_\textrm{n}$ and Probe \#4, 6, 8, 11, 17, 29, 30, 32, 35 for $B_\textrm{t}$.}
    \label{fig:imp_example}
\end{figure}

\begin{table}[]
\centering
\caption{The imputation results shown in Figure \ref{fig:imp_example} with KSTAR shot \#20341 at 2.1 sec.}
\label{tab:imputation}
\resizebox{0.48\textwidth}{!}{%
\begin{tabular}{crrcrr}
\hline \hline
\multicolumn{3}{c}{$B{_n}$ {[}T{]} $\times 10^{-2}$} & \multicolumn{3}{c}{$B{_t}$ {[}T{]} $\times 10^{-2}$} \\ 
\hline
No. & \multicolumn{1}{c}{Measured} & \multicolumn{1}{c}{Inferred} & No. & \multicolumn{1}{c}{Measured} & \multicolumn{1}{c}{Inferred} \\ \hline
3 & -1.45 & -1.88$\pm$0.22 & 4 & -14.69 & -13.97$\pm$0.47 \\
4 & -1.72 & -2.31$\pm$0.24 & 6 & -12.38 & -11.42$\pm$0.97 \\
6 & 4.62 & 4.45$\pm$0.65 & 8 & -7.82 & -7.88$\pm$0.67 \\
14 & 6.13 & 6.36$\pm$0.27 & 11 & -3.15 & -3.22$\pm$0.65 \\
18 & -8.27 & -8.11$\pm$0.48 & 17 & 0.10 & 0.30$\pm$0.52 \\
24 & 1.86 & 1.65$\pm$0.30 & 29 & 3.84 & 2.65$\pm$0.64 \\
30 & -7.52 & -7.19$\pm$0.18 & 30 & 1.15 & 0.49$\pm$0.61 \\
35 & -7.93 & -7.08$\pm$0.65 & 32 & -2.65 & -2.11$\pm$0.62 \\
37 & -4.27 & -1.41$\pm$0.93 & 35 & -8.07 & -8.87$\pm$0.55 \\ \hline \hline
\end{tabular}%
}
\end{table}

\begin{figure}
    \centering
    \includegraphics[width=0.8\linewidth]{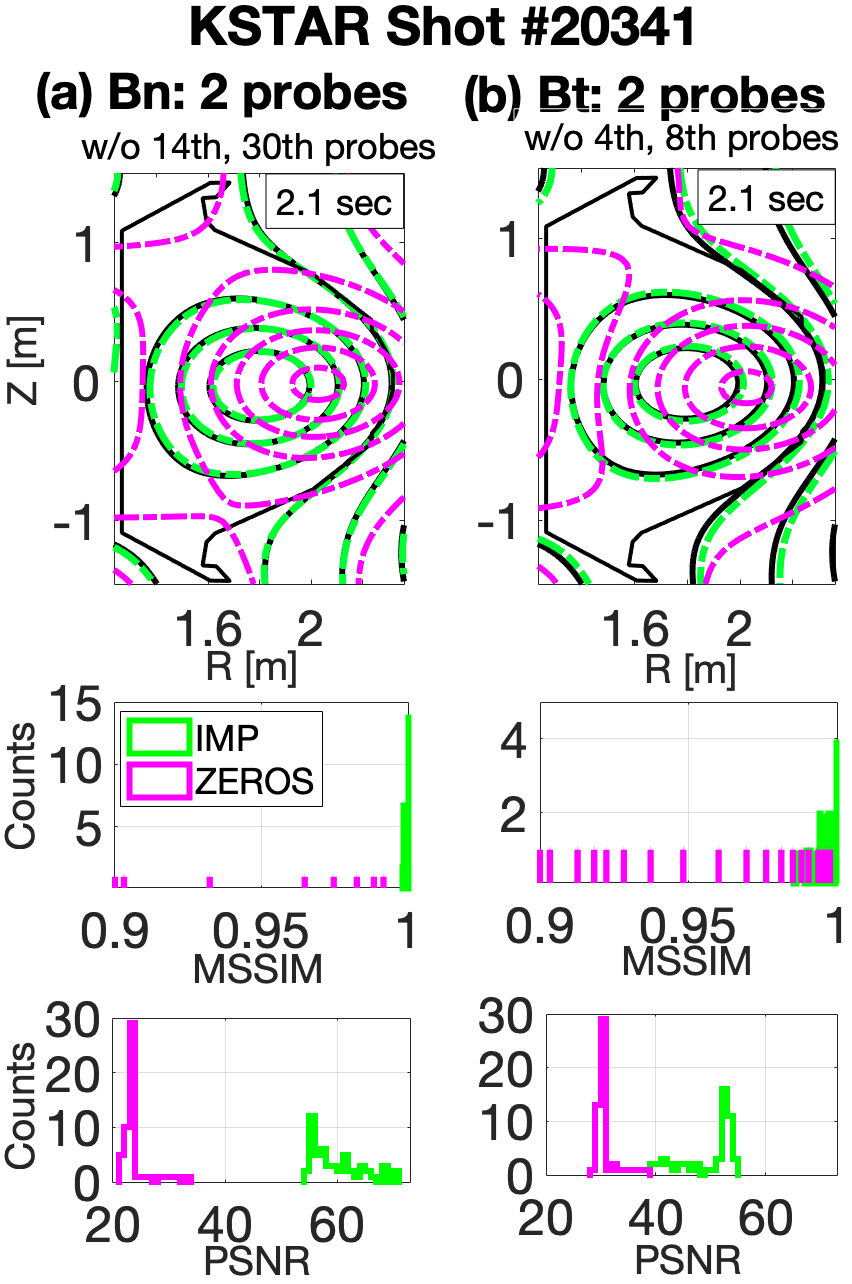}
    \caption{Top panel: nn-EFIT (NN$_{2017, 2018}$ network) reconstructed equilibria without any missing values (black line), and with two missing values replaced with the inferred values using the imputation method (green line) or with the zeros using the zero-padding method (pink dashed line), where the missing values are (a) $B_\textrm{n}$ Probe \#$14$ and $30$ (left panel) and (b) $B_\textrm{t}$ Probe \#$4$ and $8$ (right panel). Bottom panels: histograms of MSSIM and PSNR using the imputation method (green) and the zero-padding method (pink) for all the equilibria obtained from KSTAR shot \#20341\correction{, where the reference values are those obtained using nn-EFIT without any missing values}. Note that there are many more counts less than $0.9$ for MSSIM with the zero-padding method. }
    \label{fig:imp_two}
\end{figure}

We have applied the imputation method to KSTAR shot \#$20341$ at $2.1$ sec for the normal ($B_\textrm{n}$) and tangential ($B_\textrm{t}$) components of the magnetic pick-up coils as an example. We have randomly chosen nine signals from the $32$ $B_\textrm{n}$ measurements and another nine from the $36$ $B_\textrm{t}$ measurements and pretended that all of them ($9+9$) are missing simultaneously. \reffig{fig:imp_example} shows the measured (blue open circles) and the inferred (red crosses with their uncertainties) values for (a) $B_\textrm{n}$ and (b) $B_\textrm{t}$.  Probe \# on the horizontal axis is used as an identification index of the magnetic pick-up coils. \reftable{tab:imputation} provides the actual values of the measured and inferred ones for better comparisons. We find that the imputation method infers the correct (measured) values very well except Probe \#$37$ of $B_\textrm{n}$. Inferred (missing) probes are Probe \#3, 4, 6, 14, 18, 24, 30, 35, 37 for $B_\textrm{n}$ and Probe \#4, 6, 8, 11, 17, 29, 30, 32, 35 for $B_\textrm{t}$. Here, we provide all the Probe \#'s used for the neural network: $B_\textrm{n}$ Probe \#[2, $\dots$, 6, 8, 9, 11, $\dots$, 15, 17, $\dots$, 20, 23, $\dots$, 26, 28, $\dots$, 32, 34, 35, 37, $\dots$, 41] (a total of $32$) and $B_\textrm{t}$ Probe \#[2, $\dots$, 6, 8, 9, 11, $\dots$, 32, 34, 35, 37, $\dots$, 41] (a total of $36$).

Comparisons between the nn-EFIT without any missing values, which we treat as reference values, and the nn-EFIT with the imputation method or with the zero-padding method are made. Here, nn-EFIT results are obtained using the NN$_{2017, 2018}$ network. Top panel of \reffig{fig:imp_two} shows $\psi\lp R, Z\rp$ obtained from the nn-EFIT without any missing values (black line) and from the nn-EFIT with the two missing values replaced with the inferred values (green line), i.e., imputation method, or with zeros (pink dashed line), i.e., zero-padding method for (a) $B_\textrm{n}$ (left panel) and (b) $B_\textrm{t}$ (right panel) at $2.1$ sec of KSTAR shot \#20341. Probe \#14 and 30 for $B_\textrm{n}$ and Probe \#4 and 8 for $B_\textrm{t}$ are treated as the missing ones. Bottom panels compare histograms of MSSIM and PSNR using the imputation method (green) and the zero-padding method (pink) for all the equilibria obtained from KSTAR shot \#20341.

It is clear that nn-EFIT with the imputation method (green line) is not only much better than that with the zero-padding method (pink dashed line) but it also reconstructs the equilibrium close to the reference (black). In fact, the zero-padding method is too far off from the reference (black line) to be relied on for plasma controls.

\begin{figure}
    \centering
    \includegraphics[width=0.8\linewidth]{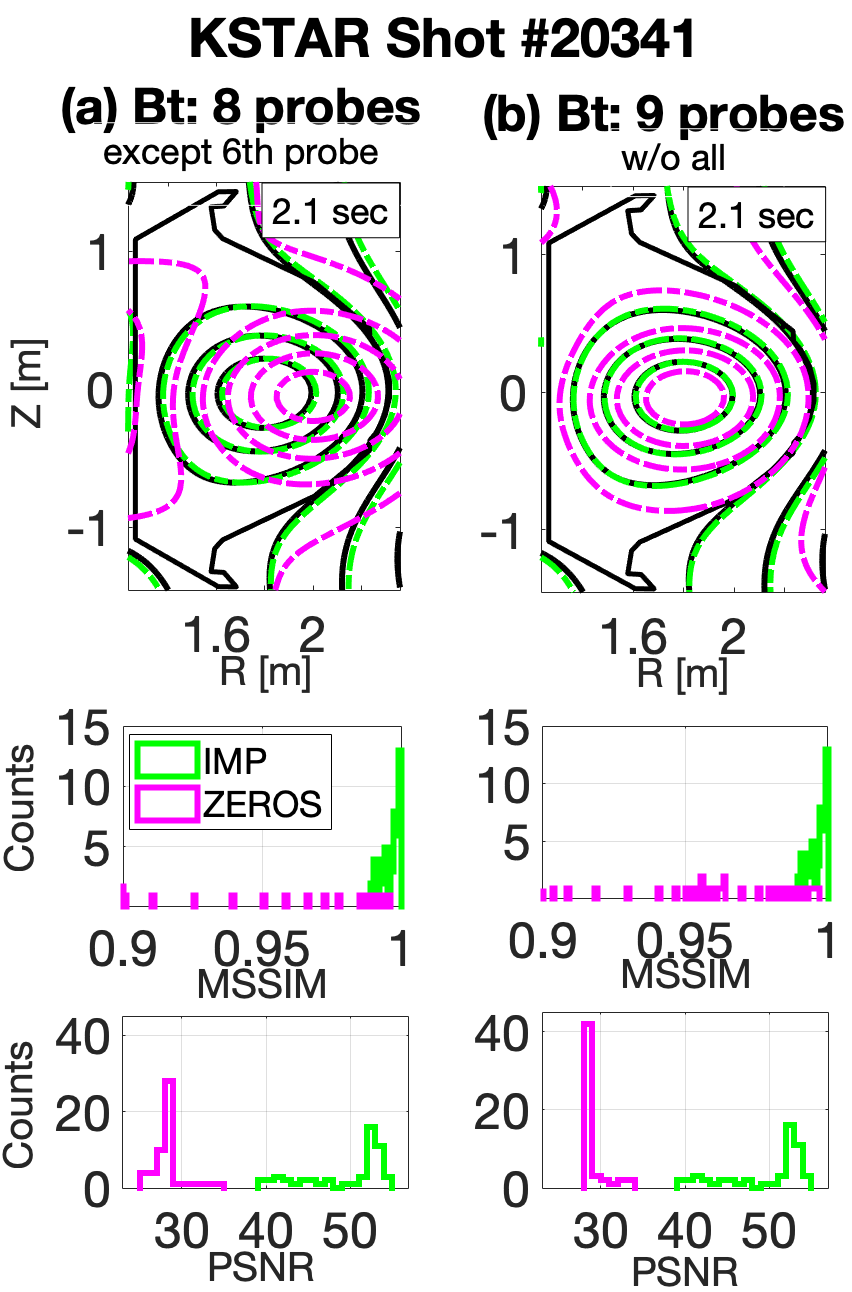}
    \caption{Same color code as in \reffig{fig:imp_two}. Missing values are (a) eight $B_\textrm{t}$ (except only Probe \#$6$) and (b) all nine $B_\textrm{t}$.}
    \label{fig:imp_many_Bt}
\end{figure}

\begin{figure}
    \centering
    \includegraphics[width=0.8\linewidth]{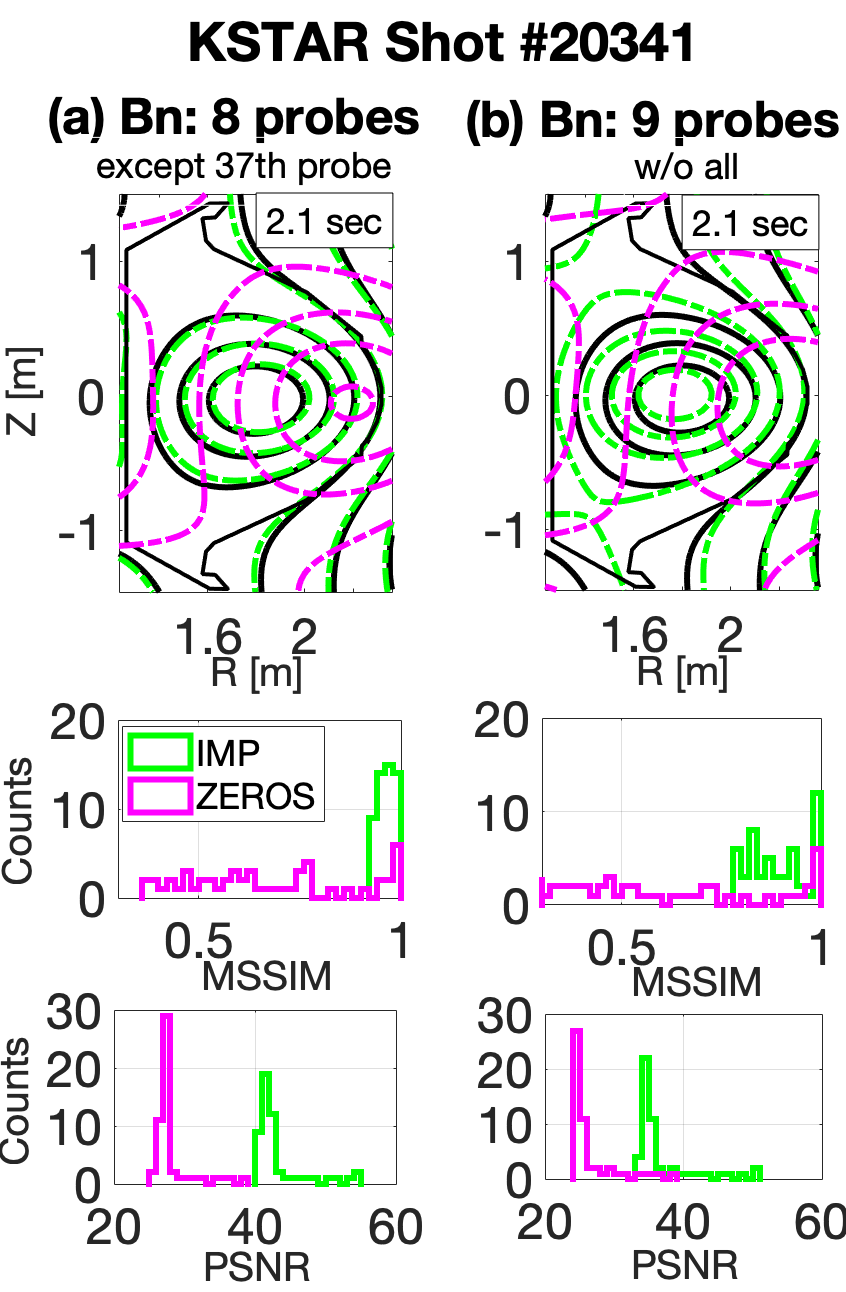}
    \caption{Same color code as in \reffig{fig:imp_two}. Missing values are (a) eight $B_\textrm{n}$ (except only Probe \#$37)$, (b) all nine $B_\textrm{n}$.}
    \label{fig:imp_many_Bn}
\end{figure}

Motivated by such a successful result of the nn-EFIT with the imputation method on the two missing values, we have increased number of missing values as shown in Figures \ref{fig:imp_many_Bt} and \ref{fig:imp_many_Bn} for the same KSTAR discharge, i.e., KSTAR shot \#20341. Let us first discuss \reffig{fig:imp_many_Bt} which are with (a) the eight (except only Probe \#6) and (b) nine (all) missing values of $B_\textrm{t}$. Color codes are same as in \reffig{fig:imp_two}, i.e., the reference is black, and nn-EFIT with the imputation method green or with the zero-padding method pink. It is evident that the nn-EFIT with the imputation method performs well at least up to nine missing values. Such a result is, in fact, expected since the imputation method has inferred the missing values well as shown in \reffig{fig:imp_example}(b) in addition to the fact that a well-trained neural network typically has a reasonable degree of resistance on noises. Again, the nn-EFIT with the zero-padding method is not reliable.

\begin{figure}
    \centering
    \includegraphics[width=0.8\linewidth]{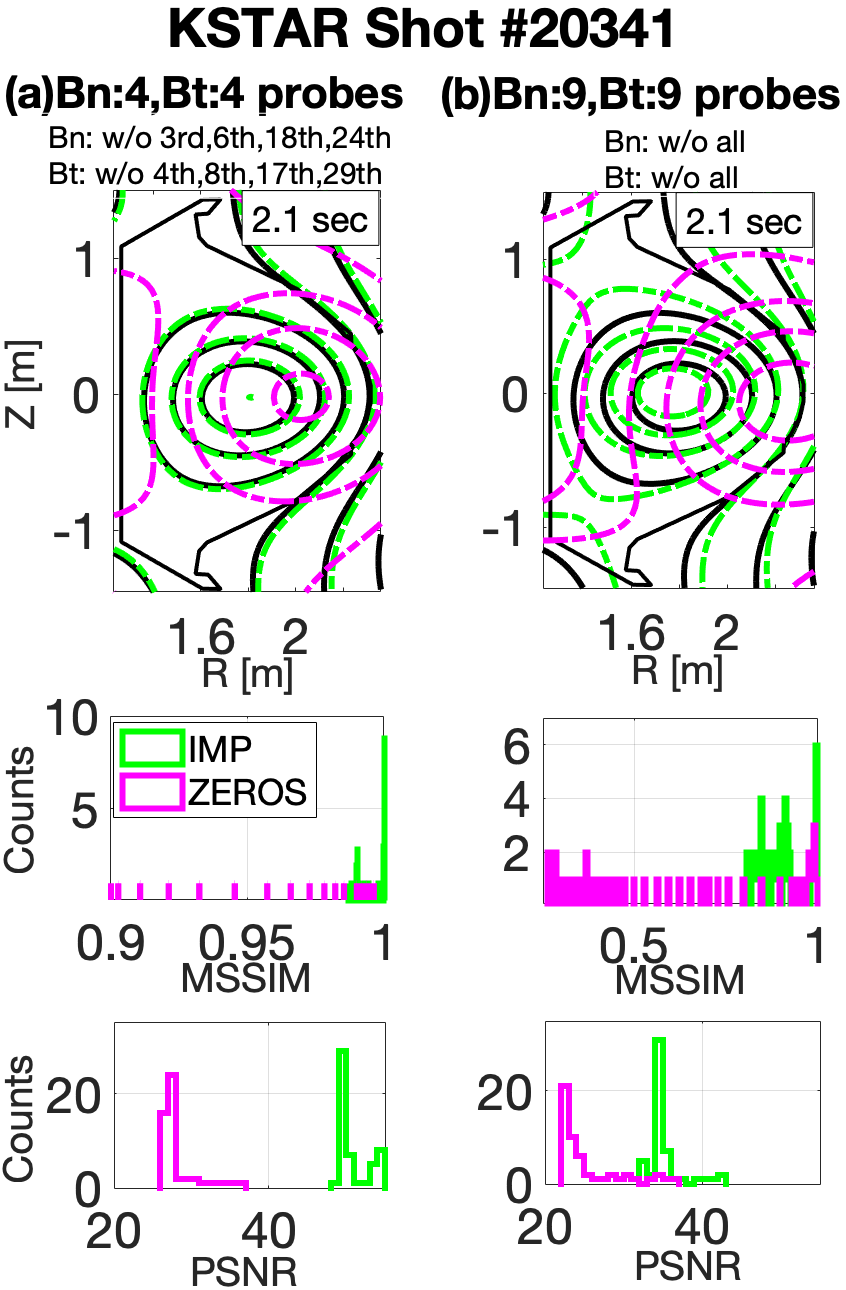}
    \caption{Same color code as in \reffig{fig:imp_two}. Combinations of missing $B_\textrm{n}$ and $B_\textrm{t}$ are examined: (a) four missing $B_\textrm{n}$ and four mssing $B_\textrm{t}$ case and (b) nine missing $B_\textrm{n}$ and nine missing $B_\textrm{t}$ case.}
    \label{fig:imp_many_comb}
\end{figure}

\reffig{fig:imp_many_Bn} (a) and (b) are results with the eight (except only Probe \#37) and nine (all) missing values of $B_\textrm{n}$, respectively. Color codes are same as in \reffig{fig:imp_two}. We find that the nn-EFIT with the eight missing values reconstructs the equilibrium similar to the reference one, while the reconstruction quality becomes notably worse for nine missing values. This is caused mostly due to poor inference of Probe \#$37$ by the imputation method (see \reffig{fig:imp_example}(a)). Nevertheless, the result is still better than the zero-padding method. \reffig{fig:imp_many_comb} shows the reconstruction results with the same color codes as in \reffig{fig:imp_two} when we have (a) $4+4$ and (b) $9+9$ combinations of $B_\textrm{n}$ and $B_\textrm{t}$ missing values simultaneously.

All these results suggest that the nn-EFIT with the imputation method reconstructs equilibria reasonably well except when the imputation infers the true value poorly, e.g., $B_\textrm{n}$ Probe \#37 in \reffig{fig:imp_example}(a) and \reftable{tab:imputation}. In fact, the suggested imputation method \cite{Joung:2018ju} infers the missing values based on the neighboring intact values (using Gaussian processes) while satisfying the Maxwell's equations (using Bayesian probability theory). Consequently, such a method becomes less accurate if (1) the neighboring channels are also missing \textit{AND} (2) the true values change fast from the neighboring values. In fact, $B_\textrm{n}$ Probe \#37 happens to satisfy these two conditions, i.e., Probe \#35 is also missing, and the true values of Probe \#35, \#37 and \#38 are changing fast as one can discern from \reffig{fig:imp_example}(a).

\section{Conclusions} \label{S6:con}

We have developed and presented the neural network based Grad-Shafranov solver constrained with the measured magnetic signals. The networks take the plasma current from a Rogowski coil, 32 normal and 36 tangential components of the magnetic fields from the magnetic pick-up coils, 22 poloidal fluxes from the flux loops, and $\lp R, Z\rp$ position of the interest as inputs. With three fully connected hidden layers consisting of 61 nodes each layer, the network outputs a value of poloidal flux $\psi$. We set the cost function used to train the networks to be a function of not only the poloidal flux $\psi$ but also the Grad-Shafranov equation $\Delta^*\psi$ itself. The networks are trained and validated with $1,118$ KSTAR discharges from 2017 and 2018 campaigns.

Treating the off-line EFIT results as accurate magnetic equilibria to train the networks, our networks fully reconstruct magnetic equilibria, not limited to obtaining selected information such as positions of magnetic axis, X-points or plasma boundaries, more similar to the off-line EFIT results than the rt-EFIT is to the off-line EFIT. Owing to the fact that $\lp R, Z\rp$ position is a part of the input, our networks have adjustable spatial resolution within the first wall. The imputation method supports the networks to obtain the nn-EFIT results even if there exist a few missing inputs. 

As all the necessary computation time is approximately $1$ msec, the networks have potential to be used for real-time plasma control. In addition, the networks can be used to provide large number of automated EFIT results fast for many other data analyses requiring magnetic equilibria.

\section*{Acknowledgement}
This research is supported by National R\&D Program through the National Research Foundation of Korea (NRF) funded by the Ministry of Science and ICT (grant numbers NRF-2017M1A7A1A01015892 and NRF-2017R1C1B2006248) and the KUSTAR-KAIST Institute, KAIST, Korea. 


\appendix

\section{Real-time preprocess on magnetic signals}
\label{app:drfit_adjust}

As shown in \reffig{fig:inputDrift} and discussed in \refsec{S2:collection}, normal ($B_\textrm{n}$) and tangential ($B_\textrm{t}$) components of magnetic fields measured by the magnetic pick-up coils and poloidal magnetic fluxes ($\Psi_\textrm{FL}$) measured by the flux loops tend to have residual drifts after calibrating the magnetic diagnostics (MDs). We train the neural networks with preprocessed, i.e., drift adjusted, magnetic signals. Therefore, we must be able to preprocess the signals in real time as well. Here, we introduce how we preprocess the magnetic signals in detail. The same preprocess is applied to all the training, validation and test data sets. Note that we do not claim that how we adjust the magnetic signals \textit{corrects} the drifts completely.

\subsection{Real-time drift adjustment with information obtained during the initial magnetization stage}

To adjust the signal drifts, we deem {\it a priori} that the signals drift linearly in time \cite{Strait:1997ds, Xia:2015cv, Ka:2008jg}. Of course, non-linear drift may well exist in the signals. However, we need to come up with a very simple and fast solution to adjust the drifts in real time with the limited amount of information. One can consider such linearization in time as taking up to the first order of Taylor expanded drifting signals. Therefore, we take the drifting components of the signals ($y_i^m$) from various types (the magnetic pick-up coils or the flux loops) of MDs to follow:
\begin{equation} \label{eq:lineardrift}
y_i^m = a_i^m t + b_i^m,
\end{equation}
where $t$ is the time. $a_i^m$ and $b_i^m$ are the slope and the offset, respectively, of a drift signal for the $i^\textrm{th}$ magnetic sensor of a type $m$ (magnetic pick-up coils or flux loops). Then, our goal simply becomes finding $a_i^m$ and $b_i^m$ for all $i$ and $m$ of interests before a plasma starts or the blip time ($t=0$) so that $y_i^m$ can be subtracted from the measured magnetic signals in real-time, i.e., preprocessing the magnetic signals for the neural networks.

 \begin{figure}
    \centerline{\includegraphics[width=0.5\textwidth]{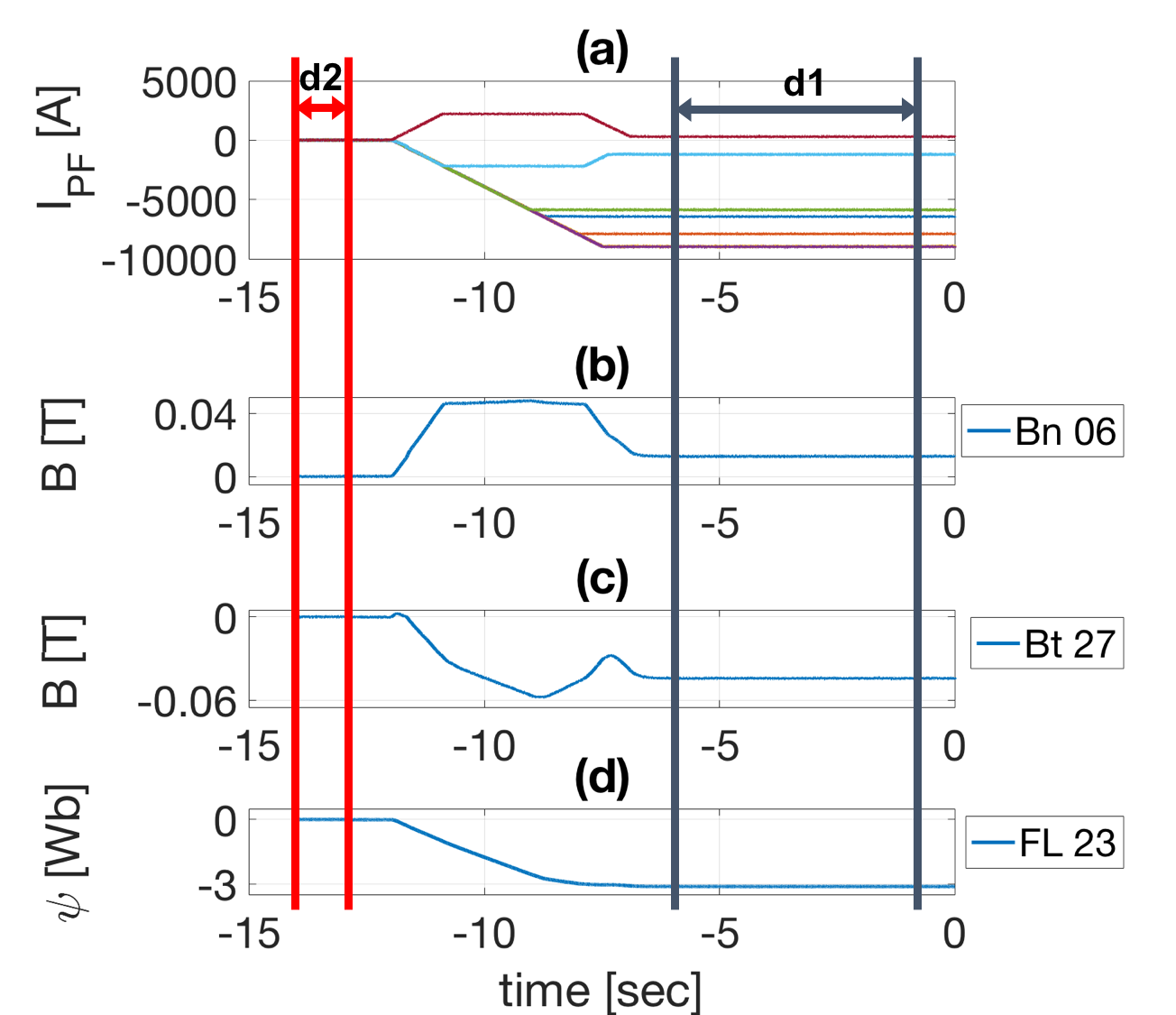}}
    \caption{An example of temporal evolutions of (a) currents in the PF coils, (b) normal and (c) tangential components of magnetic fields measured by the magnetic pick-up coils, respectively, and (d) poloidal flux measured by one of the flux loops during the initial magnetization stage, i.e., $t<0$, for a typical KSTAR discharge. Information from the time interval d1 (d2) is used to estimate $a_i^m$ ($b_i^m$).} 
    \label{PF-fig}
\end{figure}

We use two different time intervals during the initial magnetization stage, i.e., before the blip time, for every plasma discharge to find $a_i^m$ and $b_i^m$, sequentially. \reffig{PF-fig} shows an example of temporal evolutions of currents in the poloidal field (PF) coils, $B_\textrm{n}$ and $B_\textrm{t}$ and poloidal magnetic flux up to the blip time ($t=0$) of a typical KSTAR discharge.

During the time interval d1 in \reffig{PF-fig}, all the magnetic signals must be constant in time because there are no changes in currents of all the PF coils as well as there are no plasmas yet that can change the magnetic signals. Therefore, any temporal changes in a magnetic signal during d1 can be regarded as due to a non-zero $a_i^m$. With the knowledge of $a_i^m$ from d1 time interval, we obtain the value of $b_i^m$ using the fact that all the magnetic signals must be zeros during the time interval d2 because there are no sources of magnetic fields, i.e., all the currents in the PF coils are zeros. 

Summarizing our procedure, (1) we first obtain the slopes $a_i^m$ based on the fact that all the magnetic signals must be constant in time during d1 time interval, and then (2) find the offsets $b_i^m$ based on the fact that all the magnetic signals, after the linear drifts in time are removed based on the knowledge of $a_i^m$, must be zeros during d2 time interval.

\subsection{Bayesian inference}

Bayesian probability theory \cite{Sivia:2006} has a general form of
\begin{equation}
p\lp\mathcal{W}|\mathcal{D}\rp=\frac{p\lp\mathcal{D}|\mathcal{W}\rp p\lp\mathcal{W}\rp}{p\lp\mathcal{D}\rp},
\label{eq:Bayes}
\end{equation}
where $\mathcal{W}$ is a (set of) parameter(s) we wish to infer, i.e., $a_i^m$ and $b_i^m$ for our case, and $\mathcal{D}$ is the measured data, i.e., measured magnetic signals during the time intervals of d1 and d2 in Fig. \ref{PF-fig}. The \textit{posterior} $p\lp\mathcal{W}|\mathcal{D}\rp$ provides us probability of having a certain value for $\mathcal{W}$ given the measured data $\mathcal{D}$ which is proportional to a product of \textit{likelihood} $p\lp\mathcal{D}|\mathcal{W}\rp$ and \textit{prior} $p\lp\mathcal{W}\rp$. Then, we use the \textit{maximum a posterior} (MAP) to select the value of $\mathcal{W}$. The \textit{evidence} $p\lp\mathcal{D}\rp$ (or marginalized \textit{likelihood}) is typically used for a model selection and is irrelevant here as we are only interested in estimating the parameter $\mathcal{W}$, i.e., $a_i^m$ and $b_i^m$.

We estimate values of the slope $a_i^m$ and the offset $b_i^m$ based on \refeq{eq:Bayes} in two steps as described above: 
\begin{equation} \label{eq:baye-slope}
\text{Step (1)}\: : \: p(a_i^{m}|\mathcal{\vec D}_{i, d1}^m) \propto p(\vec D_{i, d1}^m|a_i^{m})p(a_i^{m}),
\end{equation}
\begin{equation} \label{eq:baye-off}
\text{Step (2)}\: : \: p(b_i^{m}|\mathcal{\vec D}_{i, d2}^m, a_i^{m*}) \propto p(\vec D_{i, d2}^m|b_i^{m}, a_i^{m*})p(b_i^{m}),
\end{equation}
where $\mathcal{\vec D}_{i, d1}^m$ ($\mathcal{\vec D}_{i, d2}^m$) are the time series data from the $i^\text{th}$ magnetic sensor of a type $m$ (magnetic pick-up coils or flux loops) during the time intervals of d1 (d2) as shown in Fig. \ref{PF-fig}. $a_i^{m*}$ is the MAP, i.e., the value of $a_i^m$ maximizing the \textit{posterior} $p(a_i^{m}|\mathcal{\vec D}_{i, d1}^m)$. Since we have no prior knowledge on $a_i^m$ and $b_i^m$, we take \textit{priors}, $p(a_i^{m})$ and $p(b_i^{m})$, to be uniform allowing all the real numbers. Note that a correct $p(a_i^{m})$ would be equal to $1/\lsb \pi\lp1+\lp a_i^{m} \rp^2 \rp \rsb$ \cite{Toussaint:RMP2011}, but we sacrifice rigor to obtain a fast solution. Furthermore, the \textit{posterior} for $b_i^m$ should, rigorously speaking, be obtained by marginalizing over all possible $a_i^m$, i.e., $p(b_i^{m}|\mathcal{\vec D}_{i, d2}^m)=\int p(b_i^{m}|\mathcal{\vec D}_{i, d2}^m, a_i^m) p(a_i^{m}|\mathcal{\vec D}_{i, d1}^m) da_i^m$. Again, as we are interested in real-time application, such a step is simplified just to use $a_i^{m*}$.

With \refeq{eq:lineardrift}, we model \textit{likelihoods}, $p(\vec D_{i, d1}^m|a_i^{m})$ and $p(\vec D_{i, d2}^m|b_i^{m}, a_i^{m*})$, as Gaussian: 
\begin{align}
& p(\vec D_{i, d1}^{m}|a_i^m) = \frac{1}{\sqrt{(2\pi)^L} |\sigma_{i, d1}^{m}| } \cr
&\! \times\! \exp \! \left(\!- \frac{\sum\limits_{t_l \in d1}^L \left[ a_i^m(t_{l} - t_0) - \left( D_{i, d1}^m(t_l) - \left< D_{i, d1}^m(t_0)\right> \right)\right]^2}{2(\sigma_{i, d1}^m)^2} ) \! \right), \cr
\label{eq:like-slope}
\end{align}
\begin{align}
p(\vec D_{i, d2}^m | b_i^m, & a_i^{m*}) = \frac{1}{\sqrt{(2\pi)^K} |\sigma_{i, d2}^m| } \cr
& \times\! \exp \! \left(\!- \frac{\sum\limits_{t_k\in d2}^K \left[ b_i^m - \left( D_{i, d2}^m(t_k) -a_i^{m*} t_k\right)\right]^2}{2(\sigma_{i, d2}^m)^2} \! \right), \cr
\label{eq:like-offset}
\end{align}
which simply state that noises in the measured signals follow Gaussian distributions. Here, $\sigma_{i,d1}^m$ and $\sigma_{i,d2}^m$ are the experimentally obtained noise levels for the $i^\text{th}$ magnetic sensor of a type $m$ (magnetic pick-up coils and flux loops) during the time intervals of d1 and d2 in \reffig{PF-fig}, respectively. $t_l$ and $t_k$ define the actual time intervals of d1 and d2, i.e., $t_l \in [-6, -1]$ sec and $t_k \in [-14, -13]$ sec with $L$ and $K$ being the numbers of the data points in each time interval, respectively. $t_0$ can be any value within the d1 time interval, and we set $t_0=-2$~sec in this work. $\left< D_{i, d1}^m(t_0)\right>$, removing the offset effect to obtain only the slope, is the time averaged value of $D_{i, d1}^m(t)$ for $t\in [t_0-0.5, t_0+0.5]$~sec. We use the time averaged value to minimize the effect of the noise in $D_{i, d1}^m(t)$ at $t=t_0$.

With our choice of uniform distributions for \textit{priors} in Equations (\ref{eq:baye-slope}) and (\ref{eq:baye-off}), MAPs for $a_i^m$ and $b_i^m$, which we denote them as $a_i^{m*}$ and $b_i^{m*}$, coincide with the maximum \textit{likelihoods} which can be analytically obtained by maximizing Equations (\ref{eq:like-slope}) and (\ref{eq:like-offset}) with respect to $a_i^m$ and $b_i^m$, respectively: 
\begin{equation} \label{eq:direct-slope}
a_i^{m*} = \frac{\sum\limits_{t_l \in d1}^L \left[ \left(D_{i, d1}^m(t_l) - \left< D_{i, d1}^m(t_0)\right>\right)\left(t_l - t_0 \right) \right]}{\sum\limits_{t_l \in d1}^L  \left[ t_l - t_0 \right]^2},
\end{equation}
\begin{equation} \label{eq:direct-off}
b_i^{m*} = \frac{1}{K} \sum\limits_{t_k \in d2}^K  \left[ D_{i, d2}^m(t_k) - a_i^{m*} t_k \right].
\end{equation}
Now, we have attained simple algebraic equations based on Bayesian probability theory which can provide us values of the slope $a_i^m$ and the offset $b_i^m$ before the blip time, i.e., before $t=0$. 

Since the required information ($a_i^m$ and $b_i^m$) to adjust drifts in the magnetic signals is obtained before every discharge starts, we can preprocess the magnetic signals in real time. This is how we have adjusted the drift signals shown in \reffig{fig:inputDrift}.

\section{Image relevant figures of merit - PSNR and MSSIM}
\label{app:image_figures}

In \refsec{S4:trainefit}, we used two image relevant figures of merit, namely PSNR (peak signal-to-noise ratio) \cite{HuynhThu:2008fm, Ebrahimi:2004fz} and MSSIM (mean structural similarity) \cite{Wang:2004gj}, to examine performance of the developed neural networks. Although these figures of merit are widely used and well known, we present short descriptions of PSNR and MSSIM for the sake of readers' convenience. Notice that we treat a reconstructed magnetic equilibrium as an image whose dimension (a number of pixels) is set by the spatial grid points.

\subsection{Peak signal-to-noise ratio (PSNR)}
\label{app:psnr}

PSNR is calculated as 
\begin{equation}
\label{eq:psnr}
\textrm{PSNR}=10\times\log_{10}\left[\frac{\max\left(y^\textrm{Target}\right)^2}{\frac{1}{M}\sum_{i=1}^M\left(y_i^\textrm{Target} - y_i^\star \right)^2}\right],
\end{equation}
where $y_i$ is the value of either $\psi$ or $\Delta^*\psi$ at the $i^\textrm{th}$ position of the spatial grid (analogous to a pixel value of an image), and $M$ for the total number of the grid points, i.e., either $286 (=22\times 13)$ or $4225 (=65\times 65)$ depending on our choice for reconstructing an equilibrium. $\max(\cdot)$ operator selects the maximum value of an argument, and $y^\textrm{Target}$ is an array containing `pixel' values of a reference EFIT `image', that is a reconstructed magnetic equilibrium. $y^\star$ is also an array, and depending on whether we wish to compare the off-line EFIT result with either rt-EFIT result or nn-EFIT result, we select the corresponding values.

\subsection{Mean structural similarity (MSSIM)}
\label{app:mssim}

MSSIM is calculated as
\begin{equation}
\label{eq:mssim}
\textrm{MSSIM}=\frac{\left(2\mu_{y^\textrm{Target}} \mu_{y^\star} + C_1 \right)\left(2\sigma_{y^\textrm{Target} y^\star}+C_2 \right) }{\left(\mu^2_{y^\textrm{Target}} + \mu^2_{y^\star}+ C_1 \right)\left(\sigma^2_{y^\textrm{Target}} + \sigma^2_{y^\star}+ C_2 \right)},
\end{equation}
where $\mu_{y^\textrm{Target}}$ and $\mu_{y^\star}$ are the mean values of $y^\textrm{Target}$ and $y^\star$, respectively. Here, $y^\textrm{Target}$ and $y^\star$ mean the same as in \refsec{app:psnr}. $\sigma^2_{y^\textrm{Target}}$ and $\sigma^2_{y^\star}$ are the variances of $y^\textrm{Target}$ and $y^\star$, respectively; while $\sigma_{y^\textrm{Target} y^\star}$ is the covariance between $y^\textrm{Target}$ and $y^\star$. $C_1$ and $C_2$ are used to prevent a possible numerical instability, i.e., denominator being zero, and set to be small numbers. Following \cite{Wang:2004gj}, we have $C_1=10^{-4}$ and $C_2=9\times 10^{-4}$.

\section*{References}

\bibliographystyle{apsrev4-1}
\bibliography{jpbib}

\end{document}